\documentclass[reprint,amsmath,amssymb,aps,prd,superscriptaddress]{revtex4-2}

\usepackage{dcolumn}
\usepackage{subcaption}
\usepackage{caption}
\usepackage{hyperref}
\usepackage[pdftex]{graphicx}
\usepackage{amsthm, braket,mathrsfs}
\usepackage[T1]{fontenc}
\usepackage{orcidlink}
\usepackage{bm}
\hypersetup{colorlinks=true}

\DeclareMathOperator{\grad}{grad}
\DeclareMathOperator{\tr}{tr}
\renewcommand{\div}{{\rm div}\, }

\renewcommand{\vec}[1]{\mathbf{#1}}

\newcommand{\R}{{\mathbb{R}}}

\newcommand{\Z}{{\mathbb{Z}}}

\newcommand{\D}{{\mathscr{D}}}
\newcommand{\I}{{\mathbb{I}}}

\newcommand{\beq}{\begin{equation}}
\newcommand{\eeq}{\end{equation}}
\newcommand{\bea}{\begin{eqnarray}}
\newcommand{\eea}{\end{eqnarray}}
\newcommand{\ben}{\begin{eqnarray*}}
\newcommand{\een}{\end{eqnarray*}}
\newcommand{\eps}{\varepsilon}
\newcommand{\cd}{\partial}
\newcommand{\wt}{\widetilde}
\newcommand{\wh}{\widehat}
\newcommand{\xvec}{\vec{x}}
\newcommand{\nvec}{\vec{n}}
\newcommand{\dvec}{\vec{d}}
\newcommand{\evec}{\vec{e}}
\newcommand{\vvec}{\vec{v}}
\newcommand{\epsvec}{\bm{\eps}}
\newcommand{\ra}{\rightarrow}
\def \d{\mathrm{d}}

\newcommand{\ignore}[1]{}

\begin{document}


\preprint{APS/123-QED}
\title{Demagnetization in micromagnetics: magnetostatic self-interactions of bulk chiral magnetic skyrmions}
\author{Paul Leask\orcidlink{0000-0002-6012-0034}}\email{palea@kth.se}
\affiliation{Department of Physics, KTH Royal Institute of Technology, 10691 Stockholm, Sweden}
\author{Martin Speight\orcidlink{0000-0002-6844-9539}}\email{j.m.speight@leeds.ac.uk}
\affiliation{School of Mathematics, University of Leeds, Leeds, LS2 9JT, England, UK}

\date{\today}

\begin{abstract}
We develop a theoretical and numerical framework for three-dimensional bulk chiral magnets that includes the full magnetostatic dipole-dipole interaction and its back-reaction on the magnetization.
Assuming translational invariance along one spatial direction, we analyze the effect of dipolar interactions on three Dzyaloshinskii--Moriya interaction (DMI) terms -- Dresselhaus, Rashba, and Heusler --corresponding to Bloch, N\'eel, and antiskyrmion textures.
In the absence of the dipolar interaction, these three DMI terms are gauge-equivalent and yield degenerate skyrmion energies.
Incorporating the non-local dipole-dipole interaction breaks this degeneracy: Bloch skyrmions remain unaffected, N\'eel skyrmions shrink slightly, and Heusler antiskyrmions lose axial symmetry and stabilize into square-lattice crystals.
The system is solved using a non-local numerical relaxation method that self-consistently computes the magnetostatic potential from Poisson’s equation.
Our results show that long-range dipolar interactions can stabilize bulk antiskyrmion crystals in translationally invariant three-dimensional chiral magnets.
\end{abstract}

\maketitle



\section{Introduction}
\label{sec: Introduction}

Topological spin textures arise in magnetic systems due to the competition between the Heisenberg exchange interaction and the Dzyaloshinskii--Moriya interaction (DMI).
The Heisenberg exchange energy promotes parallel alignment of spins whereas the DMI energy favours non-collinear alignment of spins \cite{Muratov_2020}.
Magnetic skyrmions are one such type of topological spin texture.
They are topological defects in the magnetization configuration and behave as magnetic quasi-particles. 
Such vortex-like configurations were first observed in \cite{Bogdanov_1989,Bogdanov_1994} in the context of condensed matter systems and, in particular, they found skyrmions in easy-axis chiral magnetic materials with the Dzyaloshinskii--Moriya interaction.
The easy-axis, or Zeeman, interaction is essential for stability as it provides a scale for the magnetic skyrmion.

There are three different skyrmion types investigated in the literature, each associated to a particular DMI term.
The first two are the Bloch and N\'eel skyrmions, corresponding to the Dresselhaus and Rashba spin-orbit couplings (SOC), respectively.
The final type is the antiskyrmion, which appears in Heusler compounds \cite{Saha_2019,Felser_2022,Peng_2020}, so we label the associated DMI term as the Heusler DMI.
The chiral magnet can also be formulated as a $\textup{U}(1)\times\textup{SU}(2)$ gauge theory with a constant background gauge field \cite{Amari_Nitta_2023}.
Within this formulation, the three DMI terms are related by gauge transformations, which are controlled by a single rotation parameter \cite{Schroers_2020}.
The energy functional is invariant under the relevant gauge transformations, so the three skyrmion types are degenerate in energy.
An in-depth review on the current state of the theoretical and experimental realization of magnetic skyrmions in condensed matter systems can be found in \cite{Back_2020}.

The inclusion of the magnetic dipole-dipole interaction (DDI) introduces frustration into a ferromagnetic state, turning it into a frustrated spin system.
This frustration between the DDI and Heisenberg exchange interaction can lead to interesting phenomena.
The nucleation of skyrmions using an external magnetic dipole, which modifies the Zeeman interaction term, was demonstrated numerically by Garanin \textit{et al}. \cite{Garanin_2018}.
Depending on how the magnetic dipole approaches the film, a skyrmion-antiskyrmion pair can be created in the bulk or a skyrmion can be created at the edge of the film.
It was also shown that magnetic biskyrmions can be stabilized by dipolar interactions in centrosymmetric materials where the DMI term is forbidden \cite{Gobel_2019}.

Kashuba and Pokrovsky \cite{Kashuba_1993} presented a theory on the effect of long-range dipole interactions in thin ferromagnetic films.
In particular, they studied stripe domain structures in ordinary ferromagnets with no interfacial DMI.
An analytical investigation into the stabilization of magnetic skyrmions by dipolar interactions in thin films was conducted by Ezawa \cite{Ezawa_2010}.
He considered both ordinary and chiral ferromagnets, and employed the Dresselhaus DMI term in the latter, which favours Bloch skyrmions.
In both ferromagnet types, domain wall stripes and skyrmions were constructed analytically.
However, in his analysis, Ezawa considers the spin texture of a N\'eel skyrmion, which is incompatible with the Dresselhaus DMI.

A numerical study on the effect of the dipolar interaction on chiral ferromagnets was performed by Kwon \textit{at al}. \cite{Kwon_2012}.
They also considered the Dresselhaus DMI term, for which they investigated stripe domains and Bloch skyrmion lattices in the presence of an applied external magnetic field and perpendicular uniaxial anisotropy.
As the external magnetic field is increased, the microscopic spin configurations transform from a spin-spiral stripe to a skyrmion crystalline structure.
Sui and Hu \cite{Sui_2022} conducted a similar investigation (but with no anisotropies), wherein they considered the DMI term corresponding to Rashba SOC, which gives rise to N\'eel type skyrmions.
In both studies \cite{Kwon_2012,Sui_2022}, magnetic skyrmion crystals were investigated using simulated annealing with a Monte-Carlo method on small grids.
Furthermore, they both include a tuneable constant that allows them to alter the strength of the dipolar interaction, and Kwon \textit{et al}. chose to set the dipolar interaction strength comparable to the anisotropy. 
Sui and Hu \cite{Sui_2022} find that, as the DDI strength increases, the skyrmion crystal structure changes from a hexagonal arrangement to one of a square form.

Our approach differs from that of \cite{Kwon_2012,Sui_2022}.
Instead of considering a tuneable dipole interaction parameter, we focus on the effect the dipolar interaction has on the different DMI-related skyrmion types in the bulk of a chiral ferromagnet.
That is, we do not model a thin film, instead we model a three dimensional bulk system and impose translation invariance along the $z$-direction.
We consider three DMI terms and their associated skyrmion type - Dresselhaus (Bloch), Rashba (N\'eel) and Heusler (antiskyrmion).
Without the dipolar interaction, the magnetic skyrmion types are equivalent up to redefinitions of the field.
However, the dipole-dipole interaction breaks this equivalence and there is a splitting in the energies.
Furthermore, the axial symmetry is broken for the antiskyrmion (Heusler DMI), resulting in a change in shape of both isolated antiskyrmions, and their crystal structure. 


\section{The model}
\label{sec: The model}

The system we wish to model is a large three-dimensional chiral ferromagnet whose magnetization field is assumed to vary smoothly and have constant magnitude, and so can be written $\mathbf{m}(\mathbf{x})=M_s\mathbf{n}(\mathbf{x})$ where $|\mathbf{n}|=1$ and $M_s$ is a constant parameter called the magnetic saturation density.
The total energy of such a field is assumed to take the form
\begin{equation}
    E=E_{\textup{exch}}+E_{\textup{DMI}}+E_{\textup{pot}}+E_{\textup{DDI}},
\end{equation} 
where the exchange, DMI and potential energies are
\begin{align}
    E_{\textup{exch}} = \, & \frac{J}{2}\int_{\mathbb{R}^3}\textup{d}^3x\,|\textup{d}\mathbf{n}|^2,\\
    E_{\textup{DMI}} = \, & \mathcal{D} \int_{\mathbb{R}^3} \textup{d}^3x \sum_{i=1}^3 \mathbf{d}_i\cdot(\mathbf{n}\times\partial_i\mathbf{n}),\\
    E_{\textup{pot}} = \, & \int_{\mathbb{R}^3} \textup{d}^3x \left(K_m(1-n_3^2) + M_s B_{\textup{ext}}(1-n_3)\right),\label{Epotdef}
\end{align}
and $E_{\textup{DDI}}$ is the dipole-dipole interaction energy, which we will consider in detail in section \ref{sec: The magnetic potential of a dipole distribution}. The potential has two terms: an intrinsic anisotropy term, with coupling constant $K_m$ (which can be any real constant) and a Zeeman term representing the interaction of $\mathbf{m}$ with a constant external applied magnetic field $\mathbf{B}_\textup{ext}=(0,0,B_\textup{ext})$.

The ground state configuration is found by minimizing the potential energy $E_{\textup{pot}}$, which yields the vacuum configuration $\mathbf{n}_{\textup{vac}} = \mathbf{n}_{\uparrow}=(0,0,1)$.
In general, for the energy to remain finite, we must impose the vacuum boundary condition $\lim_{|\mathbf{x}|\rightarrow\infty}\mathbf{n} = \mathbf{n}_{\uparrow}$ on configurations.
Throughout we will consider configurations invariant under translation in the $x_3$ direction (the direction of the applied magnetic field), so that the domain effectively reduces to the plane $\mathbb{R}^2$.
For these translational invariant cases, with the vacuum boundary condition, there is a compactification of space $\mathbb{R}^2 \cup \{\infty\} \cong S^2$ such that the magnetization can be interpreted as a map $\mathbf{n}: S^2\rightarrow S^2$.
Associated to this map is an integer valued topological degree $N_{\textup{Sk}}\in \pi_2(S^2) = \mathbb{Z}$, which is given by
\begin{equation}
    N_{\textup{Sk}} = \int_{\mathbb{R}^2}\textup{d}^2x\, \rho(\vec{x}), \quad \rho(\vec{x}) = \frac{1}{4\pi} \mathbf{n} \cdot \left(\frac{\partial\mathbf{n}}{\partial x_1} \times \frac{\partial \mathbf{n}}{\partial x_2}\right).
\end{equation}
Following the usual convention of the literature, magnetization fields with $N_{\textup{Sk}}<0$ are referred to as skyrmions while those with for $N_{\textup{Sk}}>0$ are antiskyrmions.
The skyrmion type is determined by the choice of DMI term.
Magnetic skyrmions with $N_{\textup{Sk}}=0$, such as a chiral magnetic bag enclosing a domain wall, are referred to as skyrmionium \cite{Rybakov_2020} or a $2\pi$-vortex \cite{Bogdanov_1999}.
We will, however, not consider skyrmion bags in this article.

In addition to magnetic skyrmions having a topological degree, they also have an associated size \cite{Wang_2018}. There is some choice in how this is 
defined \cite{Adam_2016}.
For a field theory permitting topological solitons, one can introduce the root-mean-square (RMS) radius of the soliton, defined by
\begin{equation}
    R_{\textup{Sk}} = \left( \frac{1}{N_{\textup{Sk}}} \int_{\mathbb{R}^2}\textup{d}^2x\, r^2|\rho(\vec{x})| \right)^{\frac{1}{2}}.
\end{equation}
We will use this definition of skyrmion size throughout to compare what happens to the size of the skyrmion as the DDI is included for various DMI terms.

The DMI energy is determined by three constant vectors $\mathbf{d}_i$, $i=1,2,3$.
Having restricted attention to translation invariant fields, only $\mathbf{d}_1$, $\mathbf{d}_2$ are relevant.
The overall
DMI strength $\mathcal{D}$ is set by choosing the longer of these two vectors, without loss of generality $\mathbf{d}_1$, to have length $1$.
We will consider three different choices of the DMI vectors $\mathbf{d}_1,\mathbf{d}_2$.
The most standard choice, deriving from Dresselhaus spin-obit coupling, yields the DMI vectors $\{\mathbf{d}_1=-\mathbf{e}_1, \, \mathbf{d}_2=-\mathbf{e}_2\}$, and the corresponding DMI term is
\begin{equation}
\label{eq: DMI - Dresselhaus}
    \sum_{i}\mathbf{d}_i\cdot(\mathbf{n}\times\partial_i\mathbf{n}) = \mathbf{n} \cdot \left( \bm{\nabla} \times \mathbf{n} \right) = \Lambda_{xz}^{(y)}-\Lambda_{yz}^{(x)},
\end{equation}
where we have introduced the Lifshitz invariants \cite{Leonov_2016},
\begin{equation}
    \Lambda_{ij}^{(k)} = n_i \frac{\partial n_j}{\partial x_k} - n_j \frac{\partial n_i}{\partial x_k}.
\end{equation}
\begin{figure*}[t]
    \begin{center}
    \begin{subfigure}[b]{0.3\textwidth}
    \includegraphics[width=\textwidth]{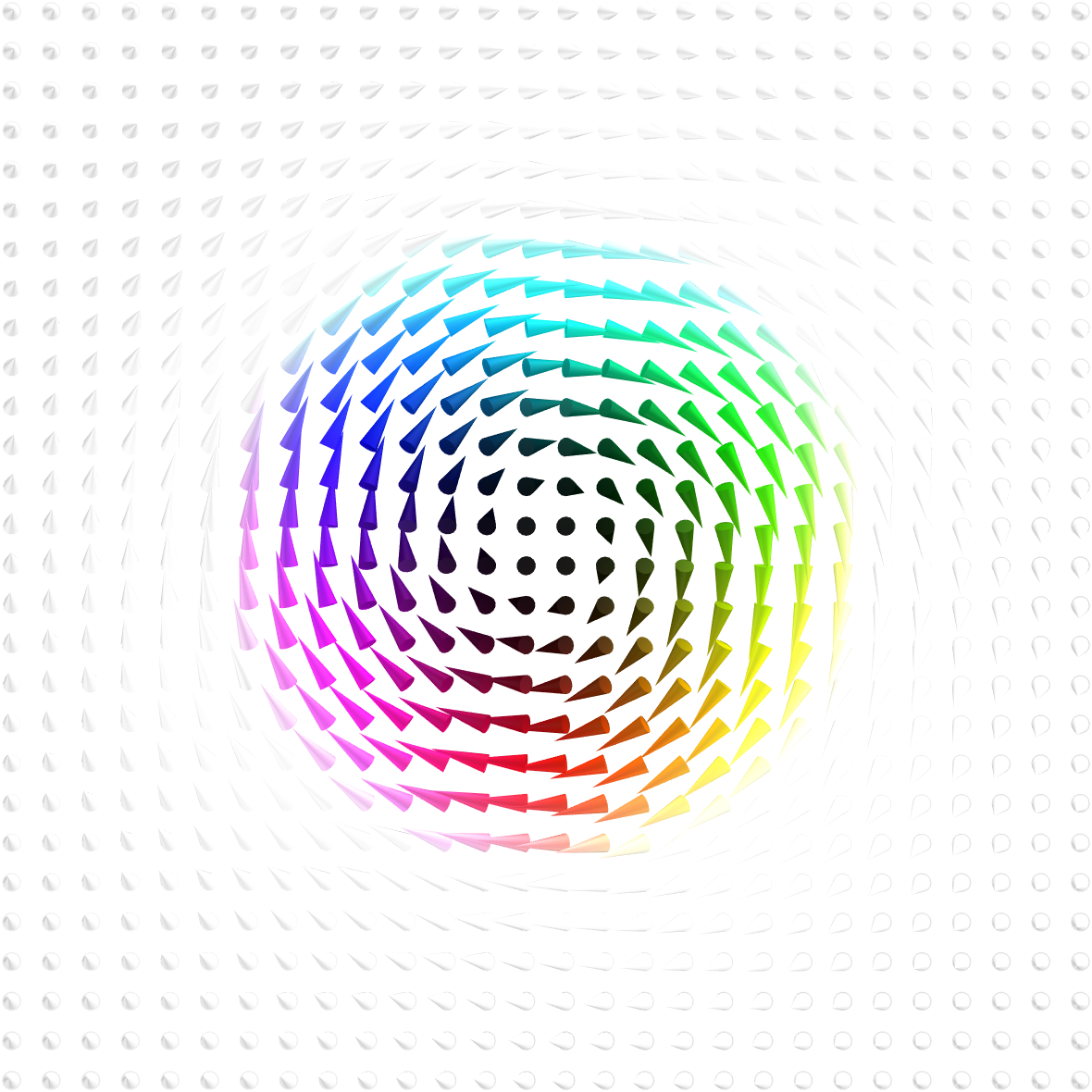}
    \caption{Bloch}
    \end{subfigure}
    ~
    \begin{subfigure}[b]{0.3\textwidth}
    \includegraphics[width=\textwidth]{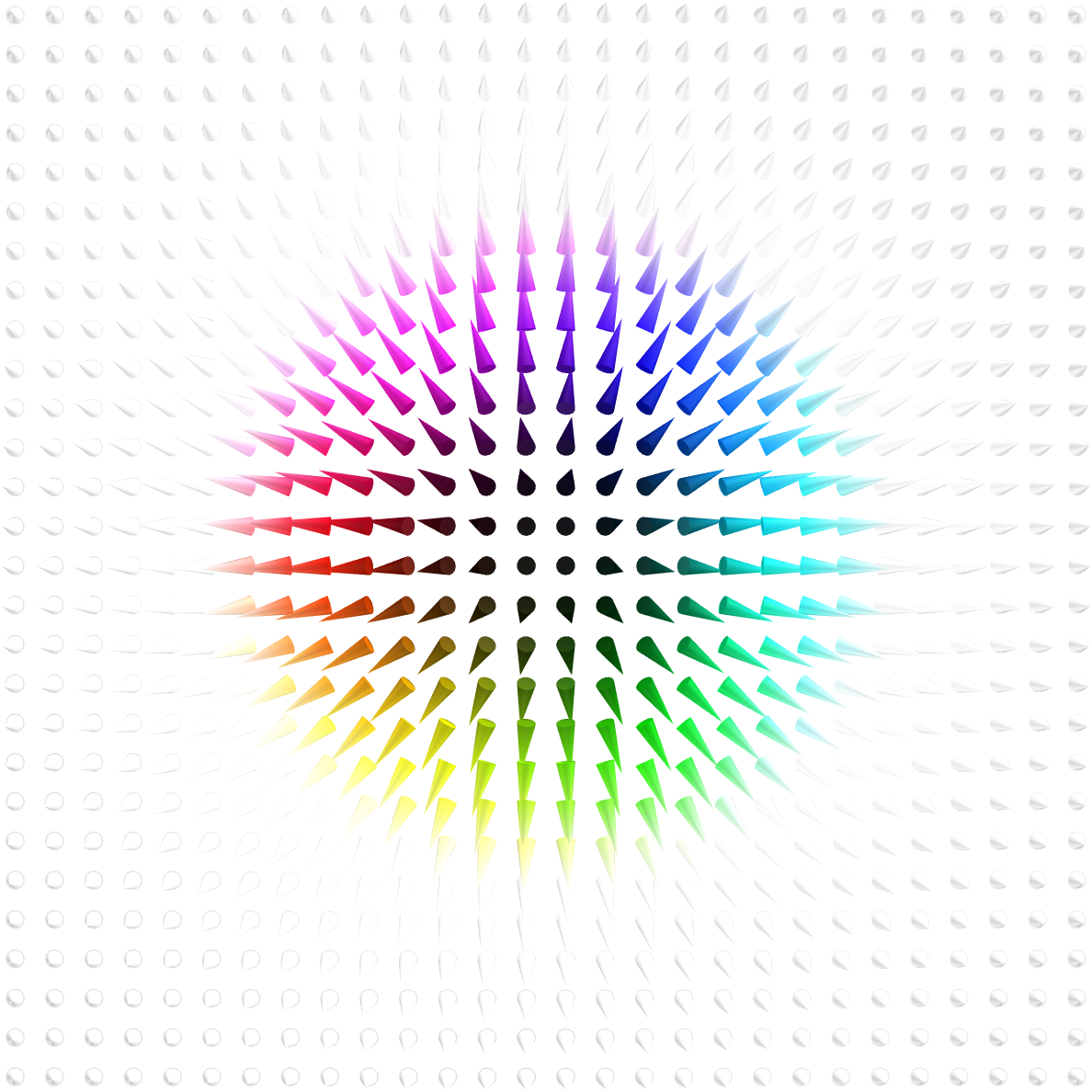}
    \caption{N\'eel}
    \end{subfigure}
        ~
    \begin{subfigure}[b]{0.3\textwidth}
    \includegraphics[width=\textwidth]{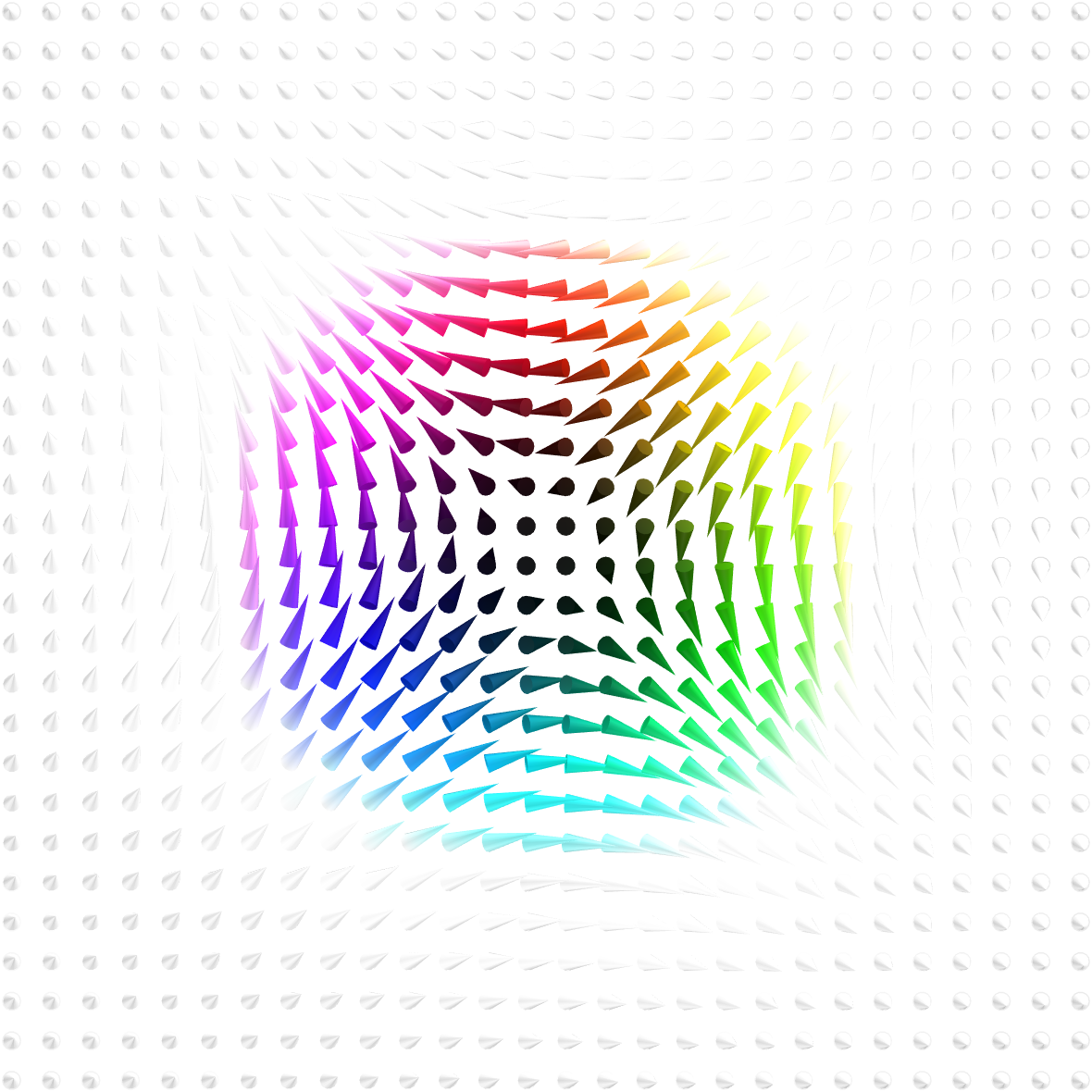}
    \caption{Antiskyrmion}
    \end{subfigure}
    \end{center}
    \caption{Three different skyrmion types consider in the text. These are the spin textures of the (a) Bloch skyrmion, (b) N\'eel skyrmion, and (c) antiskyrmion. They correspond to skyrmions with Dresselhaus, Rashba and Heusler DMI terms, respectively, and are obtained by minimizing the energy without including the magnetostatic self-interaction. The magnetization $\mathbf{n}=(n_1,n_2,n_3)\in S^2$ is coloured using the Runge colour sphere. Spin-up states $\mathbf{n}_{\uparrow}=(0,0,1)$ are white, whereas spin-down states $\mathbf{n}_{\downarrow}=(0,0,-1)$ are black. The hue is determined by the phase $\arg(n_1 + i n_2)$. Plotted on top of the skyrmion is the projection of the magnetization onto the plane $\mathbb{R}^2$, showing the in-plane vector $\left(n_1(x_1,x_2),n_2(x_1,x_2)\right)\in\mathbb{R}^2$.}
    \label{fig: Skyrmion types}
\end{figure*}
Rashba spin orbit coupling corresponds to the DMI vectors $\{\mathbf{d}_1=\mathbf{e}_2, \, \mathbf{d}_2=-\mathbf{e}_1\}$, which gives
\begin{equation}
\label{eq: DMI - Rashba}
    \sum_{i}\mathbf{d}_i\cdot(\mathbf{n}\times\partial_i\mathbf{n}) = n_3\left( \bm{\nabla}\cdot \mathbf{n}\right) - \mathbf{n}\cdot \bm{\nabla}n_3 = \Lambda_{zx}^{(x)}-\Lambda_{yz}^{(y)}.
\end{equation}
The final DMI term we consider arises from the vectors $\{\mathbf{d}_1=\mathbf{e}_1, \, \mathbf{d}_2=-\mathbf{e}_2\}$ and corresponds to tetragonal compounds with $D_{2d}$ symmetry (such as Heusler compounds) \cite{Nayak_2017},
\begin{equation}
\label{eq: DMI - Heusler}
    \sum_{i}\mathbf{d}_i\cdot(\mathbf{n}\times\partial_i\mathbf{n}) = \Lambda_{xz}^{(y)}+\Lambda_{yz}^{(x)}.
\end{equation}
In the absence of $E_{\textup{DDI}}$ all these choices are equivalent up to field redefinitions. That is,
\begin{equation}\label{eh}
    \begin{split}
        E_\textup{Dresselhaus}(n_1,n_2,n_3)=E_\textup{Rashba}(n_2,-n_1,n_3),\\
        E_\textup{Heusler}(n_1,n_2,n_3)=E_\textup{Rashba}(-n_2,-n_1,n_3).
    \end{split}
\end{equation}
The Rashba DMI term favours skyrmions of N\'eel type, that is,
\begin{equation}\label{eq: Neel ansatz}
    \mathbf{n}_{\textup{N\'eel}}(r,\theta)=(\sin f(r)\cos\theta,\sin f(r)\sin\theta,\cos f(r)),
\end{equation}
where $f(r)$ decreases monotonically from $\pi$ to $0$.
Note that this field has degree $N_{\textup{Sk}}=-1$, so we call it a skyrmion, following the standard convention in the magnetism literature.
It follows from \eqref{eh} that the Dresselhaus DMI term favours Bloch skyrmions
\begin{equation}\label{eq: Bloch ansatz}
    \mathbf{n}_{\textup{Bloch}}(r,\theta)=(-\sin f(r)\sin\theta,\sin f(r)\cos\theta,\cos f(r)),
\end{equation}
while the Heusler DMI term favours {\em antiskyrmions} of the form
\begin{equation}\label{eq: Heusler ansatz}
    \mathbf{n}_{\textup{Heusler}}(r,\theta)=(-\sin f(r)\sin\theta,-\sin f(r)\cos\theta,\cos f(r)),
\end{equation}
and that these three solutions are exactly degenerate in energy.
As we will see, dipole-dipole interactions break this degeneracy, producing particularly striking changes to the Heusler system.
A summary of the different DMI terms and their corresponding skyrmion type is provided later in Table~\ref{tab: DMI terms}, and the three different skyrmion types are displayed in Fig.~\ref{fig: Skyrmion types} with the colour scheme detailed below. 

The magnetization vector $\mathbf{n}\in S^2$ can be represented graphically with a HSV colour scheme associated to the Runge colour sphere, and is presented as laid out in \cite{Leask_2022}.
The phase $\arg(n_1 + i n_2)$ determines the hue where $\arg(n_1 + i n_2)=0$ is identified with red, $\arg(n_1 + i n_2)=2\pi/3$ with green and $\arg(n_1 + i n_2)=4\pi/3$ with blue \cite{Barsanti_2020}.
Finally, the value of $n_3$ determines the brightness \cite{Rybakov_2019}, such that spin-up states $\mathbf{n}_{\uparrow}=(0,0,1)$ are white and the spin-down states $\mathbf{n}_{\downarrow}=(0,0,-1)$ are black.

\begin{table*}[t]
    \centering
    \begin{tabular}{|c||c|c|c|}
        \hline
         & \textbf{Dresselhaus} & \textbf{Rashba} & \textbf{Heusler} \\
         \hline
         \hline
         $\{\mathbf{d}_1, \mathbf{d}_2\}$ & $\{-\mathbf{e}_1, \mathbf{e}_2\}$ & $\{\mathbf{e}_2, -\mathbf{e}_1\}$ & $\{\mathbf{e}_1, -\mathbf{e}_2\}$  \\
         \hline
         DMI & $\Lambda_{xz}^{(y)}-\Lambda_{yz}^{(x)}$ & $\Lambda_{zx}^{(x)}-\Lambda_{yz}^{(y)}$ & $\Lambda_{xz}^{(y)}+\Lambda_{yz}^{(x)}$  \\
         \hline
         Skyrmion & Bloch & N\'eel & Antiskyrmion  \\
         \hline
         $N_{\textup{Sk}}$ & $-1$ & $-1$ & $+1$  \\
         \hline
         $\mathbf{n}$ & $\begin{pmatrix}
         -\sin f(r) \sin\theta \\ 
         \sin f(r) \cos\theta \\
         \cos f(r)
         \end{pmatrix}$ & $\begin{pmatrix}
         \sin f(r) \cos\theta \\ 
         \sin f(r) \sin\theta \\
         \cos f(r)
         \end{pmatrix}$ & $\begin{pmatrix}
         -\sin f(r) \sin\theta \\ 
         -\sin f(r) \cos\theta \\
         \cos f(r)
         \end{pmatrix}$ \\
         \hline
         $\bm{\nabla}\cdot\mathbf{n}$ & $0$ & $f'(r)\cos f(r) + r^{-1}\sin f(r)$ & $\left(-f'(r)\cos f(r)+r^{-1}\sin f(r)\right)\sin2\theta$ \\
         \hline\hline
    \end{tabular}
    \caption{Summary of the three different DMI terms we consider and their corresponding skyrmion type. Here, we consider only axially symmetric configurations without the dipole-dipole interaction (DDI) present. Also shown is the divergence of each skyrmion $\bm{\nabla}\cdot\mathbf{n}$, which is relevant as it is the source $\rho=-(\bm{\nabla}\cdot\mathbf{n})$ in the Poisson equation $\Delta\psi=\mu\rho$ for the magnetic potential, and it provides information about the nature of DDI. For example,  we expect the Bloch skyrmion to be unaffected by the DDI, the magnetic potential for the N\'eel skyrmion to be axially symmetric and the Heusler antiskyrmion to lose be axial symmetry.}
    \label{tab: DMI terms}
\end{table*}


\section{The interaction energy of a distribution of magnetic dipoles}
\label{sec: The magnetic potential of a dipole distribution}

In this section we define the magnetostatic self-energy, or dipole-dipole interaction energy, $E_{\textup{DDI}}$ and compute its first variation.
The magnetic field induced by an isolated dipole of moment $\mathbf{m}$ at $\mathbf{0}$ is \cite{Qin_2018}
\begin{equation}
\label{eq: Magnetic dipole}
    \mathbf{B} = - \frac{\mu_0}{4\pi r^3} \left( \mathbf{m} - 3 \frac{\mathbf{m}\cdot \mathbf{x}}{r^2}\mathbf{x}\right), 
\end{equation}
where $\mu_0$ is the permeability of free space.
This is often referred to as the \textit{demagnetizing} field and can be expressed as the gradient of a magnetic potential $\psi:\mathbb{R}^3\rightarrow\mathbb{R}$, that is,
\begin{equation}
    \mathbf{B} = -\bm{\nabla}\psi
\end{equation}
where
\begin{equation}
    \psi = -\mu_0 \mathbf{m}\cdot \bm{\nabla}\left( \frac{1}{4\pi r}\right).
\end{equation}

Consider now a continuous distribution of magnetic dipole density $\mathbf{m}: \Omega \rightarrow \mathbb{R}^3$, where $\Omega \subseteq \mathbb{R}^3$ is some domain.
The magnetic field it induces, at a point $\mathbf{x}\in\mathbb{R}^3$, is given by integrating the field induced at $\mathbf{x}$ by $\mathbf{m}(\mathbf{y})$ at $\mathbf{y}\in\Omega$ over $\mathbf{y}\in\Omega$: 
\begin{align}\label{mls}
    \begin{split}
        \mathbf{B}(\mathbf{x}) = - \frac{\mu_0}{4\pi} \int_\Omega \textup{d}^3\mathbf{y} \frac{1}{|\mathbf{x}-\mathbf{y}|^3} \left\{ \mathbf{m}(\mathbf{y}) 
 \right.\\\left.- 3 \frac{\mathbf{m}(\mathbf{y}) \cdot (\mathbf{x}-\mathbf{y})}{|\mathbf{x}-\mathbf{y}|^2} (\mathbf{x}-\mathbf{y}) \right\}.
    \end{split}
\end{align}
Again, this is the gradient of the magnetic potential $\psi: \mathbb{R}^3 \rightarrow \mathbb{R}$,
\begin{equation}
\label{eq: Magnetic potential definition}
    \psi(\mathbf{x}) = -\mu_0 \int_\Omega \textup{d}^3\mathbf{y} \, \left\{\mathbf{m}(\mathbf{y}) \cdot \bm{\nabla}_x \left( \frac{1}{4\pi|\mathbf{x}-\mathbf{y}|} \right) \right\}.
\end{equation}

We now note that
\begin{equation}
    G(\mathbf{x},\mathbf{y})=\frac{1}{4\pi|\mathbf{x}-\mathbf{y}|}
\end{equation}
is the Green's function for the Laplacian $\Delta = -\nabla^2$ on $\mathbb{R}^3$, that is
\begin{equation}
    \Delta_x G(\mathbf{x},\mathbf{y}) = \delta(\mathbf{x}-\mathbf{y}).
\end{equation}
Comparing with \eqref{eq: Magnetic potential definition}, we see that the magnetic potential $\psi$ satisfies the PDE \cite{Fratta_2020}
\begin{equation}
    \Delta\psi = -\nabla^2\psi = \mu_0 \rho, \quad  \rho = -(\bm{\nabla}\cdot\mathbf{m}),
\end{equation}
that is, Poisson's equation with source $\rho=-(\bm{\nabla}\cdot\mathbf{m})$. Up to a multiplicative constant, the magnetic field induced by the dipole distribution $\mathbf{m}$ coincides, therefore, with the magnetic field induced by the charge distribution $-(\bm{\nabla}\cdot\mathbf{m})$ \cite{Mozurkewich_1979}.
Roughly speaking, we may think of $-(\bm{\nabla}\cdot\mathbf{m})$ as ``magnetic charge density''.

The interaction energy of a pair of magnetic dipoles $\mathbf{m}^{(1)}$, $\mathbf{m}^{(2)}$, is $-\mathbf{m}^{(1)}\cdot\mathbf{B}^{(2)}$, where $\mathbf{B}^{(2)}$ is the magnetic field induced by $\mathbf{m}^{(2)}$ at the position of $\mathbf{m}^{(1)}$. Hence, the total dipole-dipole interaction energy of a continuous dipole density distribution is the magnetostatic energy
\begin{equation}
    E_{\textup{DDI}} = -\frac{1}{2} \int_\Omega \textup{d}^3\mathbf{x} \, \left(\mathbf{B}(\mathbf{x}) \cdot \mathbf{m}(\mathbf{x}) \right),
\end{equation}
where $\mathbf{B}$ is the demagnetizing field \eqref{mls} induced by the magnetization $\mathbf{m}$.
(The factor of $1/2$ arises since we sum over dipole pairs.)
It is useful to rewrite this as a functional of the magnetic potential $\psi$:
\begin{align}\label{molasm}
    E_{\textup{DDI}} = \, & \frac{1}{2} \int_\Omega \textup{d}^3\mathbf{x} \, \mathbf{m} \cdot \bm{\nabla}\psi \nonumber \\
    = \, & -\frac{1}{2} \int_\Omega \textup{d}^3\mathbf{x} \, \left( \bm{\nabla} \cdot \mathbf{m} \right) \psi + \frac{1}{2} \oint_{\partial\Omega} \textup{d}\mathbf{s} \cdot \left( \psi\mathbf{m} \right) \nonumber \\
    = \, & \frac{1}{2\mu_0} \int_\Omega \textup{d}^3\mathbf{x} \, \psi \Delta\psi + \frac{1}{2} \oint_{\partial\Omega} \textup{d}\mathbf{s} \cdot \left( \psi\mathbf{m} \right),
\end{align}
by the Divergence Theorem. We will use this formula only in situations where the boundary conditions ensure that the boundary term vanishes.
In this case, $E_{\textup{DDI}}$ coincides with the ``electrostatic'' self-energy of the ``charge'' distribution $-(\bm{\nabla}\cdot\mathbf{m})$.

We now specialize to the case of immediate interest: the dipole-dipole interaction energy of a chiral ferromagnet in a translation invariant configuration. The dipole field has constant length $M_s$, so $\mathbf{m}=M_s\mathbf{n}$ where $\mathbf{n}$ is valued on the unit sphere. Further, we impose translation invariance in the direction $\mathbf{e}_3=(0,0,1)$, so $\mathbf{n}$ is independent of $x_3$. The boundary conditions require some care. We assume that some anisotropy in the system (either intrinsic or generated by an external applied magnetic field) imposes an energetic preference for the dipole orientation $\mathbf{n}=\mathbf{e}_3$, and consider fields $\mathbf{n}:\mathbb{R}^2\rightarrow S^2$ which have compact support in the sense that there exists $R_0>0$ such that, for all $r:=|(x_1,x_2)|\geq R_0$, $\mathbf{n}(x_1,x_2)=\mathbf{e}_3$. Since the field $\mathbf{n}$ is translation invariant, the total dipole interaction energy either vanishes (for example, if $\mathbf{n}$ is constant) or diverges. The energy per unit length (in the $\mathbf{e}_3$ direction) may be finite however, and this coincides with the total energy of the slab
$\Omega=\mathbb{R}^2\times[0,1]$, which we compute as the limit of the energy of the thick disk
$\Omega_R=\{\mathbf{x}\: :\: x_1^2+x_2^2\leq R^2,\: 0\leq x_3\leq 1\}$ as $R\rightarrow\infty$. Note that, in this case, the boundary term in \eqref{molasm} vanishes identically for all $R>R_0$: the flux of $\psi\mathbf{m}$ through the cylindrical wall vanishes since $\mathbf{m}$ has no normal component on this part of the boundary, and the flux through the top disk (at $x_3=1$) is exactly canceled by the flux through the bottom disk (at $x_3=0$) since $\psi\mathbf{m}$ is translation invariant. Hence
\begin{equation}
E_{\textup{DDI}}=\frac{1}{2\mu_0}\int_{\mathbb{R}^2}\textup{d}^2x\, \psi\Delta\psi,
\end{equation}
where $\Delta=-\partial_1^2-\partial_2^2$ is the Laplacian on $\mathbb{R}^2$, and $\psi$ satisfies
\begin{equation}\label{monlarsmi}
\Delta\psi=-\mu_0 M_s\left(\frac{\partial n_1}{\partial x_1}+\frac{\partial n_2}{\partial x_2}\right).
\end{equation}

Consider the large $r$ behaviour of the magnetic potential $\psi:\mathbb{R}^2\rightarrow\mathbb{R}$. Any solution of the Poisson equation $\Delta\psi=\mu_0\rho$ on the plane has a multipole expansion
\begin{equation}
\psi=-\frac{q\mu_0}{2\pi}\log r +O(r^{-1})
\end{equation}
where $q=\int_{\mathbb{R}^2}\textup{d}^2x\,\rho$ is the total ``charge''.
In general, such functions are logarithmically unbounded.
In our case
\begin{align}
    q=\int_{\mathbb{R}^2}\textup{d}^2x\,\rho = -\int_{\mathbb{R}^2}\textup{d}^2x\, (\bm{\nabla} \cdot \mathbf{m}) = 0,
\end{align}
and we conclude that $\psi$ is (at least) $1/r$ localized.

\section{Evading the Hobart--Derrick theorem}
\label{sec: Derrick scaling}

So far we have shown how to include the dipolar interaction and computed its variation with respect to the magnetization.
Now, we want to understand how the dipolar interaction transforms under a coordinate rescaling, to see the role it plays in evading the Hobart--Derrick theorem \cite{Derrick_1964}.

Let us consider an energy and length rescaling with $E=\frac{J^2}{\mathcal{D}}\hat{E}$ and $x=\frac{J}{\mathcal{D}}\hat{x}$.
Further, let us also consider the rescaling of the magnetic potential $\psi = \frac{\mathcal{D}}{M_s} \hat\psi$.
Then the dimensionless energy $\hat{E}$ is (having dropped all $\hat{}$ decorations) 
\begin{equation}
    \begin{split}
        E = \int_{\mathbb{R}^3} \left\{ \frac{1}{2}|\textup{d}\mathbf{n}|^2 + \mathbf{d}_i\cdot(\mathbf{n}\times\partial_i\mathbf{n}) + K(1-n_3^2) \right.\\\left. + h(1-n_3)  + \frac{1}{2}\mathbf{n} \cdot \bm{\nabla}\psi \right\} \textup{d}^3x,
    \end{split}
\end{equation}
where the magnetic potential satisfies the dimensionless Poisson equation
\begin{equation}
\label{eq: Adimensional Poisson equation}
    \Delta \psi = -\mu \bm{\nabla} \cdot \mathbf{n}.
\end{equation}
The dimensionless coupling constants are
\begin{equation}
    K = \frac{K_mJ}{\mathcal{D}^2}, \quad h = \frac{M_s B_\textup{ext}J}{\mathcal{D}^2}, \quad \mu = \frac{\mu_0JM_s^2}{\mathcal{D}^2}.
\end{equation}

We now return to the case of interest where we have translation invariance in the $\mathbf{e}_3$ direction.
Before, we were considering the energy per unit length in the translation invariant direction.
So, our domain was $\Omega=\mathbb{R}^2\times[0,1]$, which now becomes $\Omega'=\mathbb{R}^2\times[0,t]$ where $t=L_0^{-1}$ is the thickness of the slab under the rescaling.
To maintain adimensionality, let us consider the energy per unit thickness $E/t$.
Therefore, the adimensional energy functional we are really considering is
\begin{align}
\label{eq: Adimensional energy}
    \begin{split}
        E = \int_{\mathbb{R}^2} \left\{ \frac{1}{2}|\textup{d}\mathbf{n}|^2 + \mathbf{d}_i\cdot(\mathbf{n}\times\partial_i\mathbf{n}) + K(1-n_3^2) \right.\\\left. + h(1-n_3) + \frac{1}{2}\mathbf{n} \cdot \bm{\nabla}\psi \right\} \textup{d}^2x.
    \end{split}
\end{align}

With the dimensionless total energy \eqref{eq: Adimensional energy} in hand, we now turn to evading the Hobart--Derrick theorem \cite{Derrick_1964}.
Let us consider, for a given field $\nvec:\R^2\ra S^2$, the one parameter variation $\nvec_\lambda(\xvec):=\nvec(\lambda\xvec)$, $\lambda>0$, obtained by isotropic dilation. The ``magnetic charge density'' of $\nvec_\lambda$ is
$\rho_\lambda(x)=-\bm{\nabla}\cdot\nvec_\lambda(\xvec)=-\lambda\rho(\lambda\xvec)$, so its magnetic potential $\psi_\lambda$ satisfies
\beq 
(\Delta\psi_\lambda)(\xvec)=\mu\lambda\rho(\lambda\xvec).
\eeq 
Comparing with \eqref{eq: Adimensional Poisson equation}, we see that
$\psi_\lambda(\xvec)=\psi(\lambda\xvec)/\lambda$. Combining this with the well-known scaling behavior of the exchange, DMI and potential terms, we see that
\begin{equation}
    E(\nvec_\lambda) = E_{\textup{exch}}(\nvec) + \frac{1}{\lambda}E_{\textup{DMI}}(\nvec) + \frac{1}{\lambda^2} \left( E_{\textup{pot}}(\nvec) + E_{\textup{DDI}}(\nvec) \right).
\end{equation}
So the dipole-dipole interaction energy behaves like a potential term under the spatial scaling.

If $\nvec$ is a static solution of the model, it extremizes $E$ locally with respect to all variations, including $\nvec_\lambda$, so
\begin{equation}
\label{eq: Derrick scaling}
    \left.\frac{\textup{d}E}{\textup{d}\lambda}\right|_{\lambda=1} = E_{\textup{DMI}} + 2 \left( E_{\textup{pot}} + E_{\textup{DDI}} \right) = 0,
\end{equation}
the Derrick scaling constraint. 
Therefore, it is possible to evade the Hobart--Derrick theorem without inclusion of the anisotropy and Zeeman interactions, provided the dipolar interaction is present.

We note that the energy contribution from the DMI term must be negative semi-definite as both the potential and DDI energies are positive semi-definite.
To see this, we use the general identity $\psi\Delta\phi=\bm{\nabla}\psi\cdot\bm{\nabla}\phi - \bm{\nabla}\cdot \left( \psi\bm{\nabla}\phi \right)$ and the divergence theorem to express the DDI energy as the electrostatic self-energy of the charge distribution $\bm{\nabla}\cdot\mathbf{n}$, that is
\begin{equation}
    E_{\textup{DDI}} = \frac{1}{2\mu}\int_{\mathbb{R}^2} \textup{d}^2x \, |\bm{\nabla}\psi|^2 \geq 0.
\end{equation}
Hence, the dipolar interaction can stabilize magnetic skyrmions in chiral ferromagnets where the DMI is present, but does not provide stability in ordinary ferromagnets.


\section{The Euler-Lagrange equations}
\label{sec: EL eqns}

We seek fields $\nvec:\R^2\ra S^2$ which (at least locally) minimize $E$ so, for all smooth variations $\mathbf{n}_t$ of $\mathbf{n}=\mathbf{n}_0$ through fields of compact support, we require that
\beq
\frac{\textup{d}}{\textup{d}t}\bigg|_{t=0}E(\nvec_t)=0.
\eeq
This condition is equivalent to the Euler-Lagrange equations for $E$, which we now derive.

Let
$\epsvec=\partial_t\mathbf{n}_t|_{t=0}$, the generator of the variation $\nvec_t$, and note that $\epsvec\cdot\nvec= 0$ everywhere since $|\nvec|=1$.
The induced variations of $E_{\textup{exch}}$, $E_{\textup{DMI}}$ and
$E_{\textup{pot}}$ are easily computed:
\bea
    \frac{\textup{d}}{\textup{d}t}\bigg|_{t=0}E_{\textup{exch}}(\nvec_t)&=&\int_{\R^2} \textup{d}^2x\, \epsvec\cdot\Delta\nvec    \label{exchv}   \\
\frac{\textup{d}}{\textup{d}t}\bigg|_{t=0}E_{\textup{DMI}}(\nvec_t)&=&\int_{\R^2}\textup{d}^2x\,\epsvec\cdot(-2\dvec_i\times\cd_i\nvec)     \label{DMIv}   \\
\frac{\textup{d}}{\textup{d}t}\bigg|_{t=0}E_{\textup{pot}}(\nvec_t)&=&\int_{\R^2}\textup{d}^2x\,
\epsvec\cdot(0,0,-h-2Kn_3). \label{potv}
\eea
We will also need the variation of $E_{\textup{DDI}}$ which is rather more subtle. 

Denote by $\psi_t$ the unique solution of \eqref{eq: Adimensional Poisson equation}
with source $-\div \mathbf{n}_t$ decaying to $0$ at infinity, and $\dot\psi=\partial_t\psi_t|_{t=0}$. It is important to note that, while 
$\epsvec$ has compact support, neither $\psi=\psi_0$ nor $\dot\psi$ do: as argued above, they are $1/r$ localized. Care must be taken with boundary terms when computing the variation of $E_{\textup{DDI}}$, therefore.  The variation of $E_{\textup{DDI}}$ induced by $\mathbf{n}_t$ is
\begin{align}
    \frac{\textup{d}\: }{\textup{d}t}\bigg|_{t=0}E_{\textup{DDI}}(\mathbf{n}_t) = \, &\frac{1}{2\mu} \int_{\mathbb{R}^2} \textup{d}^2x \left(\dot\psi\Delta\psi+\psi\Delta\dot\psi\right) \nonumber \\
    = \, &\frac{1}{\mu}\int_{\mathbb{R}^2}\textup{d}^2x\,\psi\Delta\dot\psi
    \nonumber \\
    & +\lim_{R\rightarrow\infty}\frac{1}{2\mu}\int_{\partial B_R(0)}(\dot\psi*\textup{d}\psi-\psi *\textup{d}\dot\psi) \nonumber \\
    = \, &\frac{1}{\mu}\int_{\mathbb{R}^2}\textup{d}^2x\,\psi\Delta\dot\psi
    \nonumber \\
    &+\lim_{R\rightarrow\infty}\frac{R}{2\mu_0}\int_0^{2\pi}(\dot\psi\psi_r-\psi\dot\psi_r) \textup{d}\theta\nonumber \\
    = \, &\frac{1}{\mu}\int_{\mathbb{R}^2}\textup{d}^2x\, \psi\Delta\dot\psi
\end{align}
since $\psi,\dot\psi=O(r^{-1})$ and $\psi_r,\dot\psi_r=O(r^{-2})$. Differentiating \eqref{eq: Adimensional Poisson equation} with respect to $t$, we deduce that
\begin{equation}
    \Delta\dot\psi = -\mu\bm{\nabla}\cdot\bm{\eps},
\end{equation}
and hence
\begin{align}
    \frac{\textup{d}\: }{\textup{d}t}\bigg|_{t=0}E_{\textup{DDI}}(\mathbf{n}_t)= \, & -\int_{\mathbb{R}^2}\textup{d}^2x\, \psi \left(\bm{\nabla}\cdot\bm{\eps}\right) \nonumber \\
    = \, & \int_{\mathbb{R}^2}\textup{d}^2x\, {\epsvec}\cdot\bm{\nabla}\psi \nonumber \\
    & -\lim_{R\rightarrow\infty}\int_{\partial B_R(0)}\psi *\varepsilon \nonumber \\
    = \, & \int_{\mathbb{R}^2}\textup{d}^2x\,\bm{\eps}\cdot\bm{\nabla}\psi
    \label{DDIv}
\end{align}
by Stokes's Theorem, since $\varepsilon=\varepsilon_1 \textup{d} x_1+\varepsilon_2\textup{d} x_2$ has compact support. 

Combining \eqref{exchv}, \eqref{DMIv}, \eqref{potv} and \eqref{DDIv}, we see that
\bea
\frac{\textup{d}}{\textup{d}t}\bigg|_{t=0}E(\nvec_t)&=&\int_{\R^2}\textup{d}^2x\,\epsvec\cdot
(\Delta\nvec-2\dvec_i\times\cd_i\nvec\nonumber \\ &&\qquad -(h+2K\nvec\cdot\evec_3)\evec_3+\bm{\nabla}\psi),\label{Ev}
\eea
and this must vanish for arbitrary $\epsvec$ pointwise orthogonal to $\nvec$.
Hence, the Euler-Lagrange equation for $E$ is
\begin{equation}\label{eq: Euler-Lagrange equations}
    \grad E(\mathbf{n}):=P_{\mathbf{n}}(\Delta\nvec-2\dvec_i\times\cd_i\nvec-(h+2Kn_3)\evec_3+\bm{\nabla}\psi)=0,
\end{equation}
where the projector $P_{\mathbf{n}}(\mathbf{u}):=\mathbf{u}-(\mathbf{u}\cdot\mathbf{n})\mathbf{n}$ accounts for the constraint that
$|\mathbf{n}|\equiv 1$ and $\psi$ is determined by the Poisson equation
\eqref{eq: Adimensional Poisson equation}. Our notation calls attention to the fact that $\grad E$ is formally the $L^2$ gradient of $E$, that is, the direction in field space in which $E(\nvec)$ grows fastest.


\section{Numerical method}
\label{sec: Numerical method}

We now detail our method for finding magnetization fields $\mathbf{n}$ that simultaneously minimize the energy \eqref{eq: Adimensional energy} and solve Poisson's equation \eqref{eq: Adimensional Poisson equation}.
This method is implemented for NVIDIA CUDA architecture and is adapted from a similar non-local method developed to determine (nuclear) skyrmion crystals stabilized by $\omega$-mesons \cite{Leask_Harland_2024,Leask_2024}.
Similar methods have also recently been developed to study knot solitons in an extended version of the standard model of particle physics \cite{Hamada_2025} and the back-reaction of the Coulomb force in an extension of the Skyrme model \cite{Gudnason_2025}.
We follow the same methodology, separating our algorithm into two stages, detailed as follows.

\ignore{
Firstly, now that we have derived an adimensional energy, it will be convenient to express everything in index notation for our numerical algorithm.
The generalized DMI term can be expressed in index notation as 
\begin{equation}
    \mathbf{d}_i\cdot(\mathbf{n}\times\partial_i\mathbf{n}) = (\mathbf{d}_i)_j\epsilon_{jkl}n_k\partial_i n_l = \mathcal{D}_{ji}\epsilon_{jkl}n_k\partial_i n_l,
\end{equation}
where the columns of $\mathcal{D}$ are the DMI vectors $\mathbf{d}_i$.
Then, the dimensionless energy can be written as $E = \int_{\mathbb{R}^2} \textup{d}^2x\, \mathcal{E}$, where the energy density is given by
\begin{equation}
\label{eq: Adimensional energy index}
    \begin{split}
        \mathcal{E} = \frac{1}{2}\left( \partial_j n_i \right)^2 + \mathcal{D}_{ji}\epsilon_{jkl}n_k\partial_i n_l + K(1-n_3^2) \\ + h(1-n_3) + \frac{1}{2} n_i\partial_i\psi,
    \end{split}
\end{equation}
and the associated Euler--Lagrange field equations are
\begin{equation}
\label{eq: Euler-Lagrange equations}
    \frac{\delta \mathcal{E}}{\delta n_i} = -\partial_j\partial_j n_i + 2\mathcal{D}_{ba}\epsilon_{bij}\partial_a n_j - \delta_i^3 \left( 2Kn_3 + h \right) + \partial_i\psi.
\end{equation}
}

Magnetic skyrmions are local minimizers of the energy functional \eqref{eq: Adimensional energy}.
We choose to numerically relax the energy \eqref{eq: Adimensional energy} using an accelerated gradient descent based method with flow arresting criteria, known as arrested Newton flow.
It consists of a fourth order central finite difference method, yielding a discrete approximation $E_{\textup{dis}}(\mathbf{n})$ to the energy $E(\mathbf{n})$, and is performed on a meshgrid of $N^2$ lattice sites with lattice spacing $h_{\textup{latt}}$, so the field may be thought of as a point in the discretised configuration space $\mathcal{C}=(S^2)^{N^2} \subset \mathbb{R}^{3N^2}$. Hence our discrete approximant of the energy functional is a function $E_{\textup{dis}}:\mathcal{C} \rightarrow \mathbb{R}$. To find a minimum of 
$E_{\textup{dis}}$, we start at some initial field $\nvec(0)\in\mathcal{C}$ and solve Newton's equation for the potential $E_{\textup{dis}}$
\beq
\ddot\nvec(t)=-\grad E_{\textup{dis}}(\nvec(t))
\eeq
with initial velocity $\dot\nvec(0)=0$ using an appropriate time stepping scheme (we use a 4th order Runge-Kutta scheme). This curve will accelerate ``downhill'' initially, but will not converge to a minimum of $E_{\textup{dis}}$ (since the flow conserves total energy, including kinetic energy). To remedy this, after each time step, we check whether $E_{\textup{dis}}(t+\delta t)>E_{\textup{dis}}(t)$: if it is, we set $\dot\nvec(t)=0$ and restart the flow at $\nvec(t)$.
The flow is deemed to have converged when $\lVert \grad E_{\textup{dis}}(\mathbf{n}) \rVert_\infty<\varepsilon_\textup{tol}$, where $\varepsilon_\textup{tol}$ is some threshold tolerance (we use $\varepsilon_\textup{tol}=10^{-6}$).
This scheme is simple, robust, and converges much faster than conventional gradient flow. 

In practice, we compute $\grad E_{\textup{dis}}(\nvec)$ by discretizing our formula for
$\grad E$, \eqref{eq: Euler-Lagrange equations}. Of course, constructing
$\grad E_{\textup{dis}}(\nvec(t))$ requires us to compute the magnetic potential $\psi(t)$ induced by $\nvec(t)$, that is, we must solve the PDE \eqref{eq: Adimensional Poisson equation} on our numerical mesh. We achieve this by thinking of \eqref{eq: Adimensional Poisson equation} for $\nvec$ fixed as the Euler-Lagrange equation for the functional
\begin{equation}
    F(\psi) = \frac{1}{2}\lVert\textup{d}\psi\rVert_{L^2}^2 + \mu \int_{\mathbb{R}^2}\textup{d}^2x\, \psi \left( \bm{\nabla}\cdot\mathbf{n} \right)
\end{equation}
which we minimize using a conjugate gradient method. (Such methods are particularly efficient for minimizing quadratic functionals such as this.)

To summarize, the full numerical algorithm is implemented as follows:
\begin{enumerate}
    \item Perform a step of the arrested Newton flow method for the magnetization $\mathbf{n}$ using a 4th order Runge-Kutta method.
    \item Solve Poisson's equation for the magnetic potential $\psi$ using nonlinear conjugate gradient descent with the Polak-Ribiere-Polyak method.
    \item Compute the total energy of the configuration $(\mathbf{n}_n,\psi_n)$ and compare to the energy of the previous configuration $(\mathbf{n}_{n-1},\psi_{n-1})$. If the energy has increased, arrest the flow.
    \item Check the convergence criterion: $\lVert E_{\textup{dis}}(\mathbf{n}) \rVert_\infty<\varepsilon$. If the convergence criterion has been satisfied, then stop the algorithm.
    \item Repeat the process (return to step 1).
\end{enumerate}

\subsection{Landau--Lifshitz--Gilbert dynamics}
\label{subsec: LLG dynamics}

As a consistency check, we verify our numerics by comparing a selcetion of our results with those obtained using Landau--Lifshitz--Gilbert (LLG) dynamics.
The dimensionless LLG equation is
\begin{equation}
\label{eq: LLG}
    \frac{\textup{d}\vec{n}}{\textup{d}\tau} = -\frac{\gamma}{1+\alpha^2} \vec{n} \times  \left( \vec{h}_{\textup{eff}} + \alpha \vec{n} \times \vec{h}_{\textup{eff}} \right)
\end{equation}
where $\gamma$ is the gyromagnetic ratio, $\alpha$ is the Gilbert damping parameter, the effective magnetic field is
\begin{align}
    \vec{h}_{\textup{eff}} = \, & -\frac{\delta E}{\delta \vec{n}} \nonumber \\
    = \, & -\Delta\nvec + 2\dvec_i\times\cd_i\nvec +(h+2K \nvec \cdot \evec_3) \evec_3 + \vec{h}_{\textup{demag}},
\end{align}
and $\vec{h}_{\textup{demag}}= -\bm{\nabla}\psi$ is the demagnetization field.
The first term in the LLG equation \eqref{eq: LLG} represents precession and the second term damping.
To find static solutions we set $\alpha=1$ and $\gamma=1$.
We evolve this using a fourth order Runge--Kutta method.
During the evolution we determine the optimal/appropriate magnetostatic potential $\psi$ at each intermediate RK step, i.e. when we compute the slopes $\vec{k}_i$.
We find that the results are consistent, with the LLG method being significantly slower than the arrested Newton flow method (roughly $1\sim2$ orders of magnitude slower)


\section{Results}
\label{sec: Results}

\begin{figure*}[t]
    \centering
    \begin{subfigure}[b]{\textwidth}
        \includegraphics[width=\textwidth]{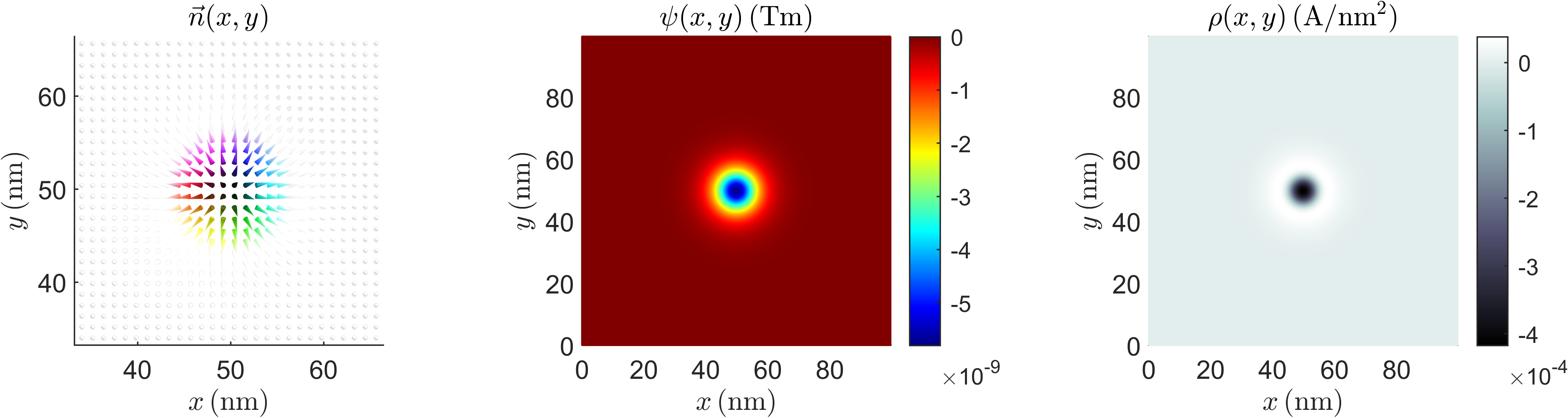}
        \caption{Rashba DMI}
    \end{subfigure}
    \\
    \begin{subfigure}[b]{\textwidth}
        \includegraphics[width=\textwidth]{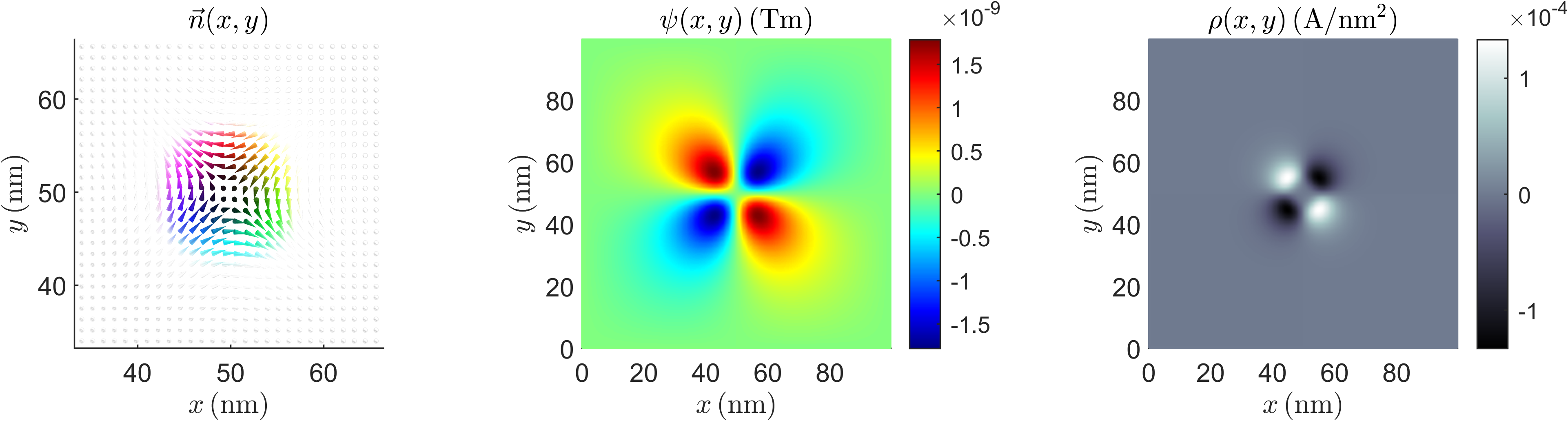}
        \caption{Heusler DMI}
    \end{subfigure}
    \caption{Effect of the dipolar interaction on the (a) N\'eel skyrmion and (b) Heusler antiskyrmion. The magnetic potential $\psi(x,y)$ is shown alongside the magnetic charge density $\rho(x,y)=-M_s(\bm{\nabla}\cdot\mathbf{n})$. The Dresselhaus DMI solution is omitted as the DDI has no effect on the Bloch skyrmion. Numerical relaxation is carried out using the arrested Netwon flow and non-linear conjugate gradient descent methods for the magnetization $\mathbf{n}:\mathbb{R}^2\rightarrow S^2$ and magnetic potential $\psi:\mathbb{R}^2\rightarrow\mathbb{R}$, respectively. We impose vacuum boundary conditions, $\mathbf{n} \rightarrow\mathbf{n}_{\uparrow}$ and $\psi\rightarrow0$, as $r\rightarrow\infty$.}
    \label{fig: DDI results - Single skyrmion}
\end{figure*}

In the previous section we detailed our numerical algorithm that constructs magnetic skyrmions in the presence of the long-range dipolar interaction, from some starting choice of initial configuration.
The method is optimized for the dimensionless energy \eqref{eq: Adimensional energy} and Poisson equation \eqref{eq: Adimensional Poisson equation} with the choice of free coupling constants $\{\mu,K,h\}$.
Although these coupling constants are dimensionless, they are defined in terms of the micromagnetic parameters $\{\mathcal{D},J,M_s,K_m,B_{\textup{ext}}\}$.

The micromagnetic parameters employed in this study are chosen to be representative of experimentally realizable chiral ferromagnets that host nanoscale skyrmions.
Specifically, we take the exchange stiffness $J=40\,\textup{pJm}^{-1}$, Dzyaloshinskii–Moriya interaction strength $\mathcal{D}=4\,\textup{mJm}^{-2}$, saturation magnetization $M_s=580\,\textup{kAm}^{-1}$, and anisotropy constant $K_m=0.8\,\textup{MJm}^{-3}$, with no external magnetic field $(B_{\textup{ext}}=0\,\textup{T})$.
This gives us the dimensionless coupling constants $K=2$, $\mu=1.057$ and $h=0$.
The chosen regime ensures a realistic competition between exchange, Dzyaloshinskii--Moriya, and dipole-dipole energies, allowing all three DMI terms to be compared on equal footing.
The inclusion of finite magnetostatic and anisotropy energies further reflects conditions relevant to both thin-film and bulk chiral magnets where dipolar effects significantly influence skyrmion stability.

In particular, these parameters correspond to the energy and length scales typical of soft ferromagnetic multilayers and Heusler-type alloys, such as CoFeB \cite{Tomasello_2014,Wang_2018} and Mn-based compounds \cite{Nayak_2017,Saha_2019}, which are widely used as benchmark materials in micromagnetic simulations of chiral spin textures.
The length scale $L_0=J/\mathcal{D}=10\,\textup{nm}$ defines the intrinsic magnetic length, yielding skyrmion diameters in the experimentally accessible range of $10-100\,\textup{nm}$.
The chosen DMI magnitude ensures that chiral twisting competes effectively with exchange and dipolar energies, allowing all three DMI terms -- Dresselhaus, Rashba, and Heusler -- to produce stable topological textures under comparable conditions.

The saturation magnetization and anisotropy values are consistent with thin-film and bulk ferromagnetic materials where demagnetizing fields are non-negligible, ensuring that the magnetostatic self-interaction plays a physically relevant role \cite{Leonov_2016}.
These values thus provide a realistic and well-balanced parameter regime for exploring how dipole-dipole interactions influence skyrmion and antiskyrmion stability in three-dimensional translationally invariant chiral magnets.

For simulations involving isolated magnetic skyrmions, all numerics carried out were simulated on a dimensionless meshgrid of size $10\times 10$ with $1024^2$ lattice sites and vacuum boundary conditions.
Physically, this corresponds to a $100\textup{nm}\times100\textup{nm}$ grid with $0.098\textup{nm}\times0.098\textup{nm}$ lattice spacing.
The energy scale for the simulations is $E_0=J^2/\mathcal{D}=4\times 10^{-19}\,\textup{J}$.


\subsection{Isolated magnetic skyrmions}
\label{subsec: Isolated magnetic skyrmions}

First of all, let us consider the effect of the dipolar interaction on a single isolated magnetic skyrmion.
Recall that we are considering three DMI terms -- Dresselhaus, Rashba and Heusler -- each of which has an associated skyrmion type (Bloch, N\'eel and antiskyrmion, respectively).
Before implementing the numerical algorithm, we can gain some intuition by computing the divergence of the magnetization ans\"atze \eqref{eq: Neel ansatz}, \eqref{eq: Bloch ansatz} and \eqref{eq: Heusler ansatz}, seeing as it is the source for the Poisson equation \eqref{eq: Adimensional Poisson equation}.
That is,
\begin{subequations}
    \begin{align}
        \bm{\nabla} \cdot \mathbf{n}_{\textup{Dresselhaus}} = \, & 0, \\
        \bm{\nabla} \cdot \mathbf{n}_{\textup{Rashba}} = \, & \frac{\textup{d}f}{\textup{d}r}\cos f(r) + \frac{1}{r}\sin f(r), \\
        \bm{\nabla} \cdot \mathbf{n}_{\textup{Heusler}} = \, & \left( \frac{1}{r}\sin f(r) - \frac{\textup{d}f}{\textup{d}r}\cos f(r) \right)\sin2\theta.
    \end{align}
\end{subequations}
So, we see that the dipolar interaction has no effect on Bloch skyrmions as the Bloch ansatz \eqref{eq: Bloch ansatz} is solenoidal. Since $E_\textup{DDI}\geq 0$ and the Bloch skyrmion
minimizes $E$ without $E_\textup{DDI}$, it follows that the Bloch skyrmion also minimizes $E$ with $E_\textup{DDI}$ in this case: the dipole-dipole interaction has no effect on isolated skyrmions in the Dresselhaus case.

However, it does have an effect on N\'eel skyrmions and antiskyrmions.
Within the N\'eel ansatz \eqref{eq: Neel ansatz} the magnetic potential is found to be radially symmetric, whereas it has a $D_2$ symmetry within the Heusler ansatz \eqref{eq: Heusler ansatz} so the DDI breaks the radial symmetry of antiksyrmions in the Heusler case.
These observations are confirmed numerically: the relaxed solutions are plotted in Fig.~\ref{fig: DDI results - Single skyrmion}.

\begin{table}[t]
    \centering
    \begin{tabular}{c|c|c|c}
        & \textbf{Dresselhaus} & \textbf{Rashba} & \textbf{Heusler} \\
        \hline
        Skyrmion & Bloch & N\'eel & Antiskyrmion \\
        \hline
        $R_{\textup{Sk}}$ & $8.422$nm & $8.415$nm & $8.418$nm \\
        \hline
    \end{tabular}
    \caption{The effect of the DDI on the skyrmion size, $R_{\textup{Sk}}$. The Bloch skyrmion is unchanged, whereas the N\'eel and anti-skyrmion shrink relative to their size without the DDI ($R_{\textup{Sk}}=8.422$nm).}
    \label{tab: Radii}
\end{table}


\subsection{Magnetic skyrmion crystals}
\label{subsec: Magnetic skyrmion crystals}

\begin{figure}[t]
    \centering
    \includegraphics[width=0.4\textwidth]{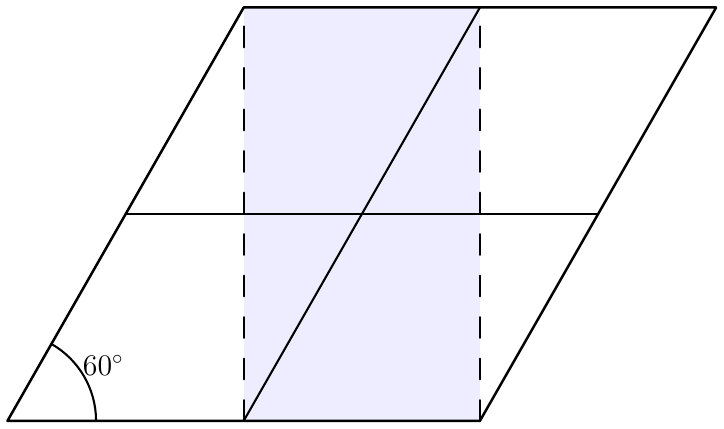}
    \caption{Fields periodic with respect to an equianharmonic lattice of side length $L$ are also periodic with respect to a rectangular lattice of size $L\times \sqrt{3} L$.}
    \label{fig:triangular}
\end{figure}

Next, we investigate the DDI effect on magnetically ordered crystals.
In the absence of DDI, there is no optimal lattice geometry for the parameter set we have chosen -- the skyrmions repel one another and want to be infinitely separated.
Our method is simpler than that of \cite{Lee_2025}: we do not vary over the lattice geometry. Instead we restrict our geometry to that of two lattice types, an equianharmonic (triangular) lattice and a lemniscatic (square) lattice. In the triangular case, for which the unit cell is a rhombus of side length $L$ and apex angle $\frac\pi3$, it is convenient to combine unit cells into pairs and work on a rectangular lattice of side lengths $L\times \sqrt 3L$, with topological charge $|N_\textup{Sk}|=2$ per unit cell, see figure \ref{fig:triangular}. For both lattice geometries, we vary the lattice parameter $L$.
In all cases, as $L\rightarrow\infty$, the energy per unit topological charge approaches that of the isolated single magnetic skyrmion.

\begin{figure*}[t]
    \centering
    \includegraphics[width=\textwidth]{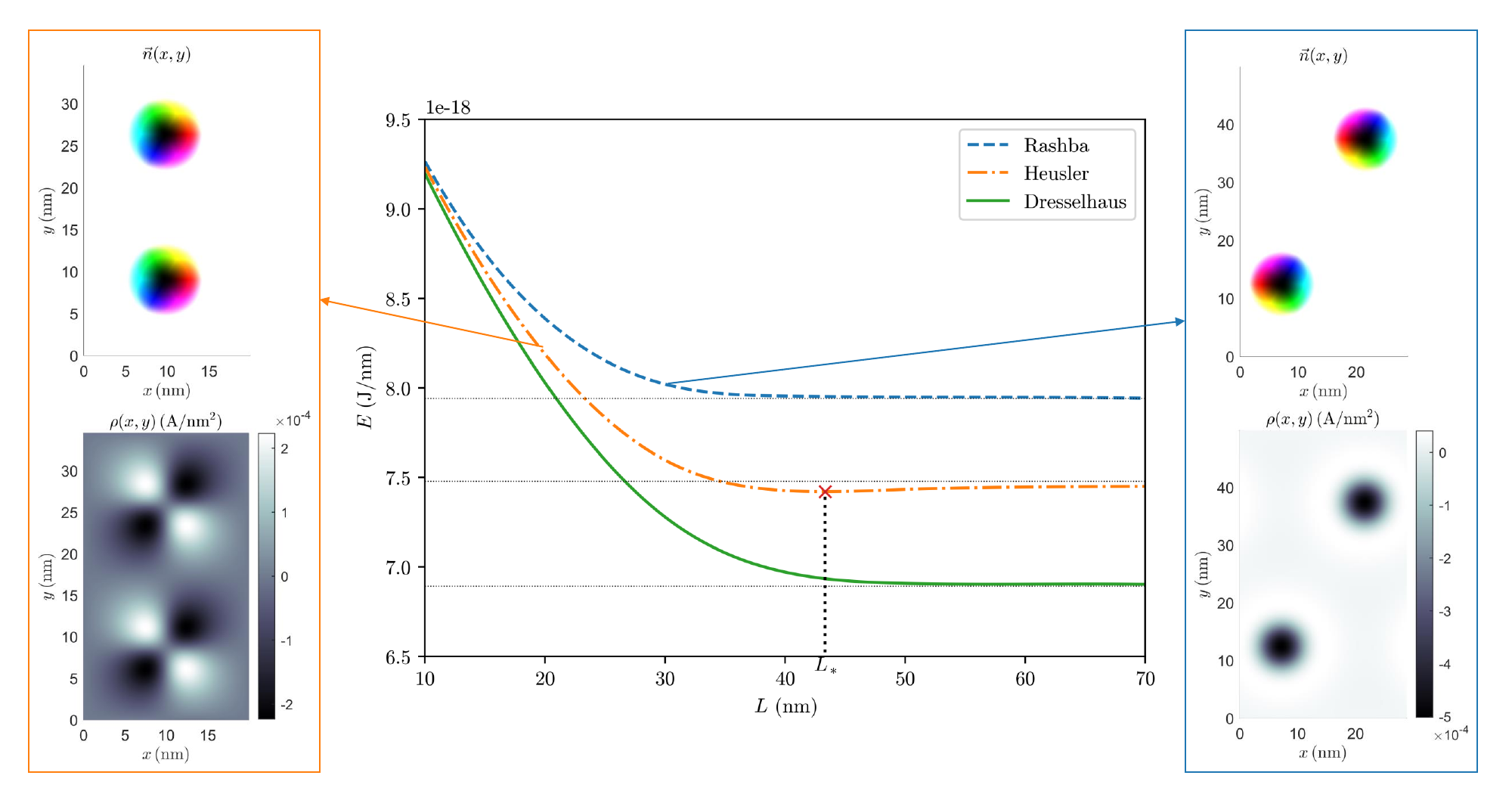}
    \caption{The effect of the dipolar interaction on hexagonal magnetic skyrmion crystals as a function of the lattice size $L$. The unit cell used is of size $L \times \sqrt{3}L$, which favours an equianharmonic lattice. Initially, two magnetic skyrmions are place in a hexagonal arrangement and then are relaxed using the numerical algorithm detailed in the text. It can be seen that including the DDI and its back-reaction lifts the energy degeneracy, resulting in distinct $E(L)$ curves. The Rashba (N\'eel) and Dresselhaus (Bloch) skyrmion lattices maintain a hexagonal arrangement, whereas the Heusler antiskyrmions relax into one of a square form. An example Heusler antiskyrmion lattice is shown at small $L=20\,\textup{nm}$ and a Rashba N\'eel skyrmion lattice is shown at a higher $L=30\,\textup{nm}$. Within these subplots is the magnetic charge density $\rho(x,y)$ and the magnetization $\mathbf{n}(x,y)$. Also plotted is the asymptotic energy of two well separated skyrmions for each DMI.}
    \label{fig: DDI results - Crystals}
\end{figure*}

This time, we include the results for the Dresselhaus DMI term.
For large lattice values $L$, the DDI will have no effect on the Bloch skyrmions.
However, for small $L$, where the individual skyrmions are packed close together, the DDI could have a small effect.
Asymptotically, in the limit $L\rightarrow\infty$, the energy per unit topological charge will approach that of two skyrmions/antiskyrmions in the absence of the DDI.

In absence of the DDI, there does not exist an optimal lattice size $L_*$ which minimizes the energy $E(L)$.
The energy is minimized for two infinitely separated skyrmions/antiskyrmions.
Consideration of the DDI breaks the energy degeneracy of the crystal configurations, resulting in separate $E$ against $L$ curves for the three DMI terms considered.
Qualitatively, the results are similar to those with no dipolar interaction for the Rashba and Dresselhaus DMIs.  This is not surprising since isolated skyrmions induce no demagnetizing field in the Dresselhaus case, and no long-range field in the Rashba case: the ``charge" distribution $\nabla\cdot\nvec$ is rotationally symmetric and has total charge $0$, so all terms in its multipole expansion vanish, and the demagnetizing field, like $\nvec$, is exponentially spatially localized.

\begin{figure*}[t]
    \centering
    \includegraphics[width=\textwidth]{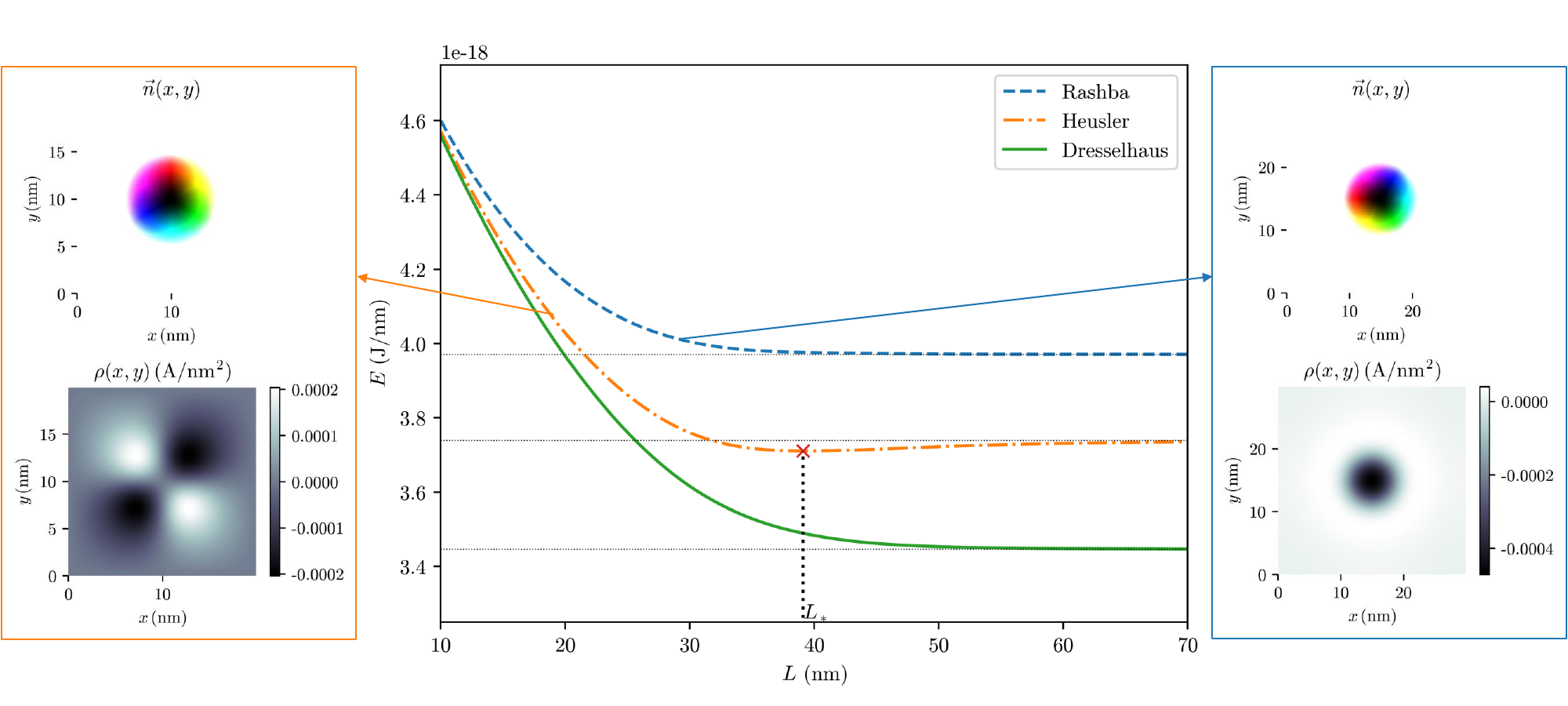}
    \caption{The effect of the dipolar interaction on square magnetic skyrmion crystals as a function of the lattice size $L$. The unit cell used is of size $L \times L$, which favours a lemniscatic lattice. Initially, two magnetic skyrmions are place in a square arrangement and then are relaxed using the numerical algorithm detailed in the text. It can be seen that including the DDI and its back-reaction lifts the energy degeneracy, resulting in distinct $E(L)$ curves. The Rashba (N\'eel) and Dresselhaus (Bloch) skyrmion lattices maintain a hexagonal arrangement, whereas the Heusler antiskyrmions relax into one of a square form. An example Heusler antiskyrmion lattice is shown at small $L=20\,\textup{nm}$ and a Rashba N\'eel skyrmion lattice is shown at a higher $L=30\,\textup{nm}$. Within these subplots is the magnetic charge density $\rho(x,y)$ and the magnetization $\mathbf{n}(x,y)$. Also plotted is the asymptotic energy of two well separated skyrmions for each DMI.}
    \label{fig: DDI results - Crystals square}
\end{figure*}

In more detail, we find that the associated skyrmions maintain their repulsive behaviour and want to be infinitely separated, but the Rashba crystals have higher energy than the Dresselhaus ones.
Another key difference is that the energy of the Rashba lattice asymptotically approaches the energy of two well-separated N\'eel skyrmions at a smaller lattice parameter $L$ than the Dresselhaus lattice does to two Bloch skyrmions.
This can be seen in Figures \ref{fig: DDI results - Crystals} and \ref{fig: DDI results - Crystals square}.
This is most likely due to the fact that the DDI slightly shrinks the size of the N\'eel skyrmion, making it more localized than the Bloch skyrmion.
Even at low lattice parameter $L$, the DDI has minimal effect on the crystalline structure of Dresselhaus Bloch skyrmions.

However, a rather striking outcome for the Heusler antiskyrmion lattice is observed.
First, the antiskyrmions prefer a square lattice arrangement over a triangular one at all lattice sizes $L$.
Furthermore, the DDI introduces an attractive force so that there does exist an optimal lattice size, $L_*=39$ nm, that minimizes the energy $E(L)$.
We have also checked that the square lattice with spacing $L_*$ minimizes energy per unit cell with respect to arbitrary variations of the {\em geometry} of the period lattice, not just its unit cell size. This calculation is quite sophisticated, and we present the details in an appendix. We are very confident, therefore, that
Heusler antiskyrmions bind together to form a square lattice, at the chosen parameter set. 

As the lattice size is increased, the energy of the Heusler lattice asymptotically approaches the energy of two infinitely separated antiskyrmions from below, meaning it has positive binding energy $E_{\textup{bind}}=2E_1-E_{\textup{latt}}>0$. By contrast, the Rashba and Dresselhaus lattices approach their respective asymptotic energies from above, that is, they have negative binding energies $2E_1<E_{\textup{latt}}$.
So, it is energetically preferable for the Heusler antiskyrmions to form a crystalline structure, whereas it is preferable for Bloch and N\'eel skyrmions to be infinitely separated.
The results of the effect of the dipolar interaction on the magnetic skyrmion crystals are summarized in Figures \ref{fig: DDI results - Crystals} and \ref{fig: DDI results - Crystals square}.

\section{Conclusion}
\label{sec: Conclusion}

We have developed a theory of magnetic skyrmions that includes the effect of the dipole-dipole interaction in the bulk of chiral ferromagnets.
In other words, we have included the demagnetizing field and its associated back-reaction on the magnetization.
The theory is formulated in terms of an arbitrary DMI and can be extended easily to include any choice of potential.
We focus on three DMI terms, these being the Dresselhaus, Rashba and Heusler, and include the Zeeman interaction and anisotropy.
In conventional chiral magnets without the DDI, all three DMI terms (and their associated skyrmion types) are equivalent up to redefinitions of the fields.
Therefore, they are energy degenerate and, for the parameter set we have chosen, do not form a stable crystalline structure.
The inclusion of the dipolar interaction, and its back-reaction, breaks the energy degeneracy and the symmetry group of the individual magnetic skyrmion.
Furthermore, it provides stability to Heusler antiskyrmions, allowing them to form into a crystalline arrangement.

The micromagnetic crystal simulation method we employed was very simple.
We restricted the geometry of our unit cell to be equianharmonic/lemniscatic, which would normally favour a hexagonal/square crystalline structure.
In both cases, the Dresselhaus and Rashba skyrmions preferred to be infinitely separated, however, the Heusler DMI preferred antiskyrmions in a square lattice arrangement.
Additionally, this remained to be the case for a different choice of parameter set, one which results in larger skyrmions of size $\sim100\textup{nm}$ and is typically used to model Heusler compounds.

An applied external magnetic field was also considered, but the results were qualitatively the same for our parameter set.
However, an external magnetic field in the absence of anisotropies was not considered in this study.
For this case, as the applied magnetic field strength is increased, one would expect to see the transition from the spin-stripe domain to a lattice of skyrmions as found in \cite{Kwon_2012}.

This method can be adapted to the analogue problem of skyrmions in chiral liquid crystals \cite{Leonov_2014}.
That is, to study the flexoelectric polarization effect on skyrmions in liquid crystals.
The problem there is slightly more difficult as the Poisson equation depends on the divergence of the polarization $\mathbf{P}(\mathbf{n})$ instead of the divergence of the order parameter $\mathbf{n}\in S^2$ \cite{Leask_2025_2}.


\appendix

\section{Lattice criticality}

We have described a numerical scheme that, for a given fixed period lattice 
\begin{equation}
\Lambda=\{\alpha_1\mathbf{v}_1+\alpha_2\mathbf{v_2}:\alpha_1,\alpha_2\in\Z\}
\end{equation}
and topological degree, produces the minimizer among all fields $\mathbf{n}:\mathbb{R}^2\rightarrow S^2$ periodic with respect to $\Lambda$
\begin{equation}
    \mathbf{n}(\mathbf{x}+\mathbf{v}_1)\equiv\mathbf{n}(\mathbf{x}+\mathbf{v}_2)\equiv\mathbf{n}(\mathbf{x})
\end{equation}
of the energy of a unit cell,
\begin{align}
    E_{\textup{cell}} = \, & \int_\Sigma \textup{d}^2x\, \left\{ \frac{1}{2}|\d\nvec|^2 + \dvec_i\cdot(\nvec\times\cd_i\nvec) + V(\nvec) \right.\nonumber \\
    & \left. +\frac{1}{2\mu}\psi\Delta\psi \right\}.
\end{align}
Here $\Sigma=\mathbb{R}^2/\Lambda\equiv T^2$ is the torus with period lattice $\Lambda$ (or, in more elementary terms, its unit cell, that is the parallelogram spanned by the vectors $\vvec_1,\vvec_2$). We have left the potential terms $V(\nvec)$ arbitrary, since our argument does not rely on the specific potential \eqref{Epotdef} used in this paper.
One should note that such a minimizer is likely to exist for {\em any} choice of period lattice $\Lambda$. If this solution is to have any physical significance, it should be a local minimizer of $E_{\textup{cell}}$ with respect to variations of the period lattice $\Lambda$ also. Previous studies of soliton crystals in other systems show that, in general, the energetically optimal period lattice may have unexpectedly low symmetry, being neither square nor triangular \cite{Speight_Winyard,Leask_2022}, so this issue is not negligible.
In this appendix, we formulate this criticality condition as a set of explicit integral constraints, suitable for numerical verification, adapting ideas introduced in \cite{Speight_2014}. 

Let $\nvec:\Sigma\ra S^2$ be a fixed field on the torus $\Sigma=\R^2/\Lambda$, assumed to minimize $E_{\textup{cell}}$ among all fields in its homotopy class. Denote by $\psi:\Sigma\ra\R$ its associated magnetic potential. Any other period lattice $\wt\Lambda$ is generated by $M\vvec_1$, $M\vvec_2$, where $M\in \textup{GL}(2,\mathbb{R})$ is some invertible matrix. Furthermore there is a canonical magnetization field on the new torus $\wt\Sigma=\R^2/\wt\Lambda$, namely $\wt{\nvec}(\xvec):=\nvec(A\xvec)$ where $A=M^{-1}$. This has an associated magnetic potential $\wt\psi:\wt\Sigma\ra\R$, which, by definition satisfies the equation
\begin{equation}
-(\cd_1^2+\cd_2^2)\wt\psi=-\mu(\cd_1\wt\nvec+\cd_2\wt\nvec_2).
\end{equation}
Note that $\wt{\psi}$ is {\em not} simply $\psi(A\xvec)$ in general.
Having fixed the mapping $\nvec$, $E_{\textup{cell}}$ is a smooth function of $A\in \textup{GL}(2,\mathbb{R})$. We require that this function is locally minimal at $A=\I_2$. In particular, $\I_2$ must be a critical point of $E_{\textup{cell}}:\textup{GL}(2,\mathbb{R})\ra\R$. 

All tori are diffeomorphic, and it is helpful to identify the new torus $\wt\Sigma$ with the reference torus $\Sigma$ via the linear diffeomorphism $M:\Sigma\ra\wt\Sigma$. From this viewpoint, the magnetization field $\nvec:\Sigma\ra S^2$ remains fixed, but the metric on $\Sigma$, its area form, the DMI vectors and the source of the magnetic potential $\wt{\psi}$ depend on $M$. The metric is the pullback of the Euclidean metric on $\wt{\Sigma}$ by $M$,
\begin{equation}
    \wt{g}=(M^T M)_{ij}\textup{d}x_i\textup{d}x_j,
\end{equation}
its area form is $\det M\,  \textup{d}x_1\wedge \textup{d}x_2$, the DMI vectors are
\begin{equation}
\wt{\dvec}_i=A_{ij}\dvec_j,
\end{equation}
and $\wh\psi:=\psi\circ L:\Sigma\ra\R$ satisfies Poisson's equation on $(\Sigma,\wt{g})$
\begin{equation}\label{sbs}
\wt{\Delta}\wh\psi=\wh\rho
\end{equation}
with source
\begin{equation}
\wh\rho=-\mu A_{ij}\cd_in_j.
\end{equation}
In equation \eqref{sbs}, 
\begin{equation}
\wt{\Delta}=-A_{ik}A_{jk}\frac{\cd^2\: }{\cd x_i\cd x_j}
\end{equation}
is the Laplacian on $\Sigma$ with respect to the deformed metric $\wt{g}$. 
One finds that
\bea
    E_{\textup{exch}}^{\textup{cell}}(A) &=& \frac{1}{2\det A}\tr(A^THA)    \\
    E_{\textup{DMI}}^{\textup{cell}}(A) &=& \frac{1}{\det A}\tr(A^T\D)   \\
    E_{\textup{pot}}^{\textup{cell}}(A) &=& \frac{1}{\det A}\int_\Sigma \textup{d}^2x\, V(\nvec) = \frac{E_{\textup{pot}}^{\textup{cell}}(\I_2)}{\det A} \\
    E_{\textup{DDI}}^{\textup{cell}}(A) &=& \frac{1}{2\mu}\int_\Sigma \frac{\textup{d}^2x}{\det A}\, \wh\psi\wt\Delta\wh\psi,
\eea
where $H$ and $\D$ are the matrices
\bea
    H_{ij} &=& \int_\Sigma \textup{d}^2x \, \cd_i\nvec\cdot\cd_j\nvec \label{Hdef}\\
    \D_{ij} &=& \int_\Sigma \textup{d}^2x \, \dvec_i\cdot(\nvec\times\cd_j\nvec).
    \label{Ddef}
\eea
Although we do not have an explicit formula for $E_{\textup{DDI}}^{\textup{cell}}(A)$, as we do for each of the other terms, we will still be able to compute its derivative at $\I_2$ explicitly, and this will suffice for our purposes. 

As a consistency check, let us derive the condition for $E_{\textup{cell}}$ to be critical with respect to isotropic dilations of the period lattice. These are generated by the curve $A(t)=e^{-t}\I_2$. The corresponding magnetic potential satisfies
\begin{equation}
e^{-2t}\Delta\wh\psi=-\mu e^{-t}\nabla\cdot\nvec,
\end{equation}
so $\wh\psi=e^t\psi$.
Hence the condition for criticality is
\bea
\frac{\textup{d}}{\textup{d}t}\bigg|_{t=0}&&\left(
\frac12\tr H+e^t\tr\D+e^{2t}E_{\textup{pot}}^{\textup{cell}}(\I_2)\right. \nonumber \\
&&\left.+\frac{1}{2\mu}\int_\Sigma \textup{d}^2x \, e^t\psi e^{-2t}\Delta e^t\psi e^{2t}\right) \nonumber \\
&=&\tr \D +2E_{\textup{pot}}^{\textup{cell}}(\I_2)
+\frac{1}{2\mu}\int_\Sigma \textup{d}^2x \, \psi\Delta\psi \nonumber \\
&=&E_{\textup{DMI}}^{\textup{cell}}(\I_2)
+2(E_{\textup{pot}}^{\textup{cell}}(\I_2)+E_{\textup{DDI}}^{\textup{cell}}(\I_2))\nonumber \\
&=&0,\label{aw}
\eea
which coincides precisely with the Derrick scaling constraint \eqref{eq: Derrick scaling} evaluated on a single lattice cell. This constraint is automatically satisfied by our numerical solutions since our scheme minimizes $E_{\textup{cell}}$ with respect to the area of the lattice cell. 

We now analyze the 3 dimensional space of lattice variations complementary to isotropic dilations. Let $A(t)$ be a curve in $\textup{SL}(2,\R)$ with $A(0)=\I_2$ generating a variation of $\Lambda$ through lattices whose unit cells have fixed area. Then $\eps:=\dot{A}(0)$ is traceless. The induced variations of the various terms in $E_{\textup{cell}}(\I_2)$ are
\bea
\frac{\textup{d}}{\textup{d}t}\bigg|_{t=0}E_{\textup{exch}}^{\textup{cell}}(A(t))
&=&\tr(\eps^T H)\\
\frac{\textup{d}}{\textup{d}t}\bigg|_{t=0}E_{\textup{DMI}}^{\textup{cell}}(A(t))
&=&\tr(\eps^T \D)\\
\frac{\textup{d}}{\textup{d}t}\bigg|_{t=0}E_{\textup{pot}}^{\textup{cell}}(A(t))
&=& 0.
\eea
The variation of $E_{\textup{DDI}}^{\textup{cell}}$ is considerably subtler. Let us denote by $\Delta_t$ the Laplacian on $\Sigma$ with respect to the metric $\wt{g}(t)$, that is,
\begin{equation}
\Delta_t=-(A(t)A(t)^T)_{ij}\cd_i\cd_j,
\end{equation}
and by $\psi_t$ the solution of
\begin{equation}\label{smbusw}
\Delta_t\psi_t=\rho_t,
\end{equation}
where
\begin{equation}
\rho_t=-\mu A_{ij}(t)\cd_in_j.
\end{equation}
Note that ${\psi}_0=\psi$. Further, let an overdot denote the derivative with respect to $t$ at $t=0$. Then
\bea
    \frac{\textup{d}}{\textup{d}t}\bigg|_{t=0}E_{\textup{DDI}}^{\textup{cell}}(A(t))
    &=& 
    \frac{1}{2\mu}\int_{\Sigma} \textup{d}^2x \left(\dot\psi\Delta\psi+\psi\Delta\dot\psi+
    \psi\dot\Delta\psi\right) \nonumber \\
    &=&\frac1\mu\int_\Sigma \textup{d}^2x \left(
    \psi\Delta\dot\psi+
    \frac12\psi\dot\Delta\psi
    \right).
\eea
Differentiating \eqref{smbusw}, we see that
\begin{equation}
\Delta\dot\psi=\dot\rho-\dot\Delta\psi
=-\mu\eps_{ij}\cd_i n_j-2\eps_{ij}\cd_i\cd_j\psi.
\end{equation}
Hence,
\bea
    \frac{\textup{d}}{\textup{d}t}\bigg|_{t=0}E_{\textup{DDI}}^{\textup{cell}}(A(t))
    &=& 
    \frac1\mu\int_\Sigma \textup{d}^2x \, \eps_{ij}\psi\left(-\mu\cd_i n_j+\cd_i\cd_j\psi\right) \nonumber \\
    &=&\tr(\eps^T Q),
\eea
where
\begin{equation}\label{Qdef}
    Q_{ij} = \int_\Sigma \textup{d}^2x \, \psi\left(\frac1\mu\cd_i\cd_j\psi-\cd_j n_i\right).
\end{equation}
Note that $Q$ is traceless by \eqref{eq: Adimensional Poisson equation}.

The condition that $E_{\textup{cell}}$ is critical with respect to all area preserving lattice variations is therefore that
\begin{equation}
\tr\left(\eps(H+\D+Q)\right)=0
\end{equation}
for all traceless matrices $\eps$. This is equivalent to the condition that
\begin{equation}\label{alwe}
H+\D+Q=\lambda \I_2
\end{equation}
for some $\lambda\in\R$. Taking traces of both sides we see that
\begin{equation}
\lambda=E_{\textup{exch}}^{\textup{cell}}+\frac12E_{\textup{DMI}}^{\textup{cell}}.
\end{equation}

To summarize, a doubly periodic map $\nvec:\R^2/\Lambda\ra S^2$, inducing magnetic potential $\psi$, is critical with respect to all variations of the period lattice $\Lambda$ if it satisfies the scaling identity 
\eqref{aw} and the angular virial constraint \eqref{alwe}, where the matrices $H,\D,Q$ are defined in \eqref{Hdef}, \eqref{Ddef}, \eqref{Qdef}, and all integrals are over a single unit cell of the lattice $\Lambda$. These conditions are easily checked numerically.

To numerically verify the extended virial constraints [\eqref{aw},\eqref{alwe}], we compute the energy normalized constraints $\{V_1=0,V_2=0\}$, where
\begin{align}
    V_1 = \, & \frac{E^{\textup{cell}}_{\textup{DMI}} + 2 \left( E^{\textup{cell}}_{\textup{pot}} + E^{\textup{cell}}_{\textup{DDI}} \right)}{E_{\textup{cell}}}, \\
    V_2 = \, & \frac{H+\D+Q-\lambda \I_2}{E_{\textup{cell}}}.
\end{align}
We find that both of these normalized constraints are satisfied to within $\textup{err}=10^{-5}$ for the square Heusler antiskyrmion lattice at a lattice parameter of $L_*=39\,\textup{nm}$.






\section*{Acknowledgments}

The authors acknowledge funding from the Olle Engkvists Stiftelse through the grant 226-0103 and the Roland Gustafssons Stiftelse för teoretisk fysik (Leask), and the UK Engineering and Physical Sciences Research Council through grant EP/Y033256/1 (Speight).


\section*{Data availability}

The code used to produce the data in this text can be found at the publicly available github repository \href{https://github.com/Paulnleask/cuDemag}{https://github.com/Paulnleask/cuDemag}.


\bibliography{main.bib}

\end{document}